\documentclass[aps,pre,nofootinbib,nobibnotes,reprint, citeautoscrip]{revtex4-2}

\usepackage{amsmath,amssymb}
\usepackage{graphicx}
\usepackage{array}
\usepackage{hyperref}
\usepackage{float}

\graphicspath{{figures/}}

\newcommand{\suppmat}{Appendix}
\usepackage[english]{babel}

\newcommand{\fs}[1]{{#1}}

\begin{document}

\title{Weighted directed clustering: interpretations and requirements for heterogeneous, inferred, and measured networks}
\author{Tanguy Fardet$^{1,2}$}
\author{Anna Levina$^{1,2}$}
\affiliation{
  $^1$ \mbox{University of T\"ubingen, T\"ubingen, Germany}\\
  $^2$ \mbox{Max Planck Institute for Biological Cybernetics, T\"ubingen, Germany}
}

\begin{abstract}
Weights and directionality of the edges carry a large part of the information we can extract from a complex network.
However, many network measures were formulated initially for undirected binary networks. 
The necessity to incorporate information about the weights led to the conception of the multiple extensions, particularly for definitions of the local clustering coefficient discussed here.
We uncover that not all of these extensions are fully-weighted; some depend on the degree and thus change a lot when an infinitely small weight edge is exchanged for the absence of an edge, a feature that is not always desirable. 
We call these methods ``hybrid'' and argue that, in many situations, one should prefer fully-weighted definitions.
After listing the necessary requirements for a method to analyze many various weighted networks properly, we propose a fully-weighted continuous clustering coefficient that satisfies all the previously proposed criteria while also being continuous with respect to vanishing weights.
We demonstrate that the behavior and meaning of the Zhang--Horvath clustering and our new continuous definition provide complementary results and significantly outperform other definitions in multiple relevant conditions.
Using synthetic and real-world examples, we show that when the network is inferred, noisy, or very heterogeneous, it is essential to use the fully-weighted clustering definitions.
\end{abstract}

\maketitle

\tableofcontents

\section{Introduction}

\fs{The clustering coefficient (CC) was originally introduced for binary undirected networks to quantify strong connectedness within a local neighborhood.}
It was defined as the fraction of all possible triangles that were realized i.e. the ratio between all triangles to which node $i$ participates ($n_{\Delta, i}$) and the total number of triangles that could theoretically be made given its degree $d_i$, which is the number of triplets ($n_{T,i}$):
\begin{equation}
	C_i^{\text{bin}} = \frac{n_{\Delta, i}}{n_{T,i}} = \frac{n_{\Delta, i}}{d_i (d_i - 1)}
\end{equation}
From a neighbor-centric perspective, it can be seen perhaps more intuitively as the probability that two neighbors of a node are connected.
However, as network science expanded, more and more graphs were encountered, where directedness and edge weights play a central role.
Generalizations of the clustering coefficient were therefore introduced to account for asymmetry in the connections between pairs of nodes or heterogeneity in their strength.

\fs{The importance of clustering, including its directed variants, to understand complex dynamics on networks has been stressed in multiple studies~\cite{Mason2007, Ahnert2008, Zhang2016, Li2018}.}
This is notably the case for the middleman motif which is a marker of feedforward loops in transcriptional networks, and of information transfer redundancy e.g. in neuroscience.
More generally, such motifs will influence the evolution of dynamical processes on the networks, for instance synchronization patterns, and have been shown to characterize families of networks such as transcription or language networks~\cite{Ahnert2008}.
Finally, clustering is used in other measurements to access the small-world propensity of networks~\cite{Muldoon2016} and the choice of a specific definition can therefore influence whether the network of interest will register as small-world or not.

\fs{In many applications network topology and weights are measured only up to certain precision~\cite{Young2021,Guimera2009}.}
For example, in neuroscience, the functional connectivity networks measured using the indirect inference of connections from the recorded activity~\cite{Fox2010Clinical,orlandi2014transfer}. 
Accepting the inevitability of noise in a network brings forward new requirements on the network measures, namely that they are stable to the noise and do not change dramatically if the weights are perturbed or weak connections are randomly omitted.

\fs{There is no agreement among the researchers which weighted extension of the clustering coefficient definition is most appropriate}. 
The three predominantly used methods at the moment~\cite{Barrat2004, Onnela2005, Zhang2005} differ in many properties of their definitions. 
Part of the reason for the absence of a single best weighted clustering lies in a different interpretation of weights in various datasets. 
Consequently, a different weighted extension might be most appropriate for various data and specific scientific questions. 
However, to understand which method to use when and why we need to understand their differences precisely.

\fs{The difficulty of extending graph measures to weighted networks is not specific to the clustering coefficient but can occur whenever ratios of degrees or path-length are involved.}
We will therefore also discuss a second clustering-related measure, called the closure coefficient and introduced as the fraction of all open walks of length 2 starting from node $i$ that are part of a triangle~\cite{Yin2019}.
This will also enable us to discuss the complementarity of closure and clustering as the former provides an important complement to analyze the tendency of nodes to form 3 and 4-cliques.

\fs{We introduce here a distinction between fully-weighted and hybrid definitions and discuss why, for several classes of networks, fully-weighted and directed definitions should be preferred to other clustering definitions that are currently used for network analysis.}
We also propose a new definition that obeys additional conditions, including continuity of the results with respect to infinitesimal changes in edge weights, which has significant consequences for the resilience to noise in inferred networks.
We demonstrate why fully-weighted methods are essential for measured and inferred networks, that are pervasive in biological fields such as neuroscience, and for networks dealing with flows of information, money, or goods that display a very broad weight distribution.

\section{Interpretation and purpose of weighted clustering}

\subsection{Desired properties of weighted clustering coefficients}

\fs{Weighted measures are crucial for many network types where the binary connectivity is either uninformative (fully connected network) or displays similar or lower heterogeneity compared to the weighted structure.}
In this study, we focus on two classes of real-world networks:  inferred or measured networks where there can be a large number of spurious (false positive) edges with small weights; and networks associated with flows of information or goods, which often display broad weight distributions.
This is notably the case for many networks in neuroscience, and more generally in information, transportation, or other social and economic networks.
Weights are essential to understand the dynamical processes that occur in these networks, requiring measures that go beyond the binary structure.

\fs{There could be multiple requirements for weighted clustering coefficients~\cite{Saramaki2007} depending on the particular question of interest and on the network properties.}
The main requirements that we considered necessary for a weighted clustering coefficient are:
\begin{itemize}
	\item \emph{normalization} ($C_i \in [0, 1]$),
	\item \emph{consistency} with the binary definition (for binary networks, it should give back the classical result),
	\item \emph{linearity} (scaling by $\alpha$ all edges involving node $i$ and all edges in triangles including node $i$ scales $C_i$ by $\alpha$),
	\item \emph{continuity} (weak influence of the addition or deletion of edges having very small weights, meaning that an edge with infinitesimally small weight should be equivalent to the absence of that edge).
\end{itemize}

Compared to a previously proposed list of conditions~\cite{Saramaki2007}, we added a \emph{continuity} condition but did not include a requirement of a specific normalization factor (the global $\max(w)$) as long as the \emph{normalization} condition is fulfilled since only the normalization matters.
We omitted the last two conditions of Saram\"aki's paper  (invariance under weight permutation and ignorance of weights not participating in any triangle).
Although they might be of interest for some specific applications, we do not consider them to be generally desired properties for a clustering coefficient.
We also did not require that all weights in a triangle should be accounted for because this condition is necessarily met if the \emph{continuity} condition is fulfilled.

\emph{Continuity} can be expressed mathematically as follow: for a graph $G(V, E)$, if a weighted edge $(u, v, w)$ with $u, v \in V$ and weight $w\in \mathbb{R}$ is added to this graph to form a new graph $G'(V, E')$, with $E' = E + \{(u, v, w)\}$, then the clustering measure is continuous if and only if $\forall i \in V$, $C_i^{(G')} \xrightarrow[w \rightarrow 0^+]{} C_i^{(G)}$.
This condition in crucial to ensure a reasonable behavior of the clustering coefficient in inferred networks.

\fs{Though some definitions of previously proposed weighted clustering coefficient definitions obey most of the required properties, none of them completely fulfill the \emph{continuity} condition --- despite previous claims \cite{Wang2016}.}
 This is why we will later propose a new definition that fulfills all aforementioned conditions.
An extensive comparison of the properties fulfilled by different clustering definitions can be found in \suppmat{} \ref{app:compare-prop}.

\subsection{State of the art for weighted clustering}

\fs{First we introduce and classify the existing weighted clustering coefficient definition.}
For all clustering definitions in the main text, we use the following notation: $A$ is an adjacency matrix, $W = \{w_{ij}\}$ is the normalized weight matrix, obtained from the original weight matrix $\tilde{W} = \{\tilde{w}_{ij}\}$ by $w_{ij} = \tilde{w}_{ij} / \max_{i,j}\left(\tilde{w}_{ij}\right)$. 

\textbf{Hybrid definitions} were the first extensions of the binary clustering coefficients definitions. 
They combine properties associated with weighted connectivity matrix (i.e. intensity of the triangle) with properties that could be already obtained from adjacency matrix (i.e. node degrees).
These definitions move from an integer counting the number of triangles ($n_\Delta$) to a sum of real numbers (computed as a function of edge weights) that we call ``intensities'' of triangles ($I_\Delta$).
The choice of a particular function for the intensity of the triangles determines the properties of the clustering coefficient.

Two popular hybrid weighted clustering were given by the teams of Barrat~\cite{Barrat2004} and Onnela~\cite{Onnela2005}.

For a node $i$ in a graph, the definition from~\cite{Barrat2004} quantifies the fraction of the node's strength that is invested in triangles (see Appendix \ref{app:barrat} for more details):
\begin{equation}
	C^B_i = \frac{\left(WA^2\right)_{ii}}{2 s_i (d_i - 1)} = C^{\text{bin}}_i \frac{\overline{w^\Delta_i}}{\overline{w_i}}
\end{equation}
where $s_i = \sum_{j \neq i} w_{ij}$ is the strength of the node, $\overline{w_i}$ is the average weight of the edges involving $i$, and $\overline{w^\Delta_i} =  \sum_{j\neq k} \frac{w_{ij} + w_{ik}}{2 n_{\Delta, i}} a_{ij}a_{ik}a_{jk} $ is the average weight of edges involving $i$ that are part of a triangle computed over all triangles to which node $i$ participates.
In terms of triangle intensity, this definition was originally written:
\begin{equation}
\begin{split}
	C^B_i	&= \frac{\sum_{j\neq k} \frac{w_{ij} + w_{ik}}{2} a_{ij}a_{ik}a_{jk}}{2 s_i (d_i - 1)}\\
			&= \frac{1}{d_i(d_i-1)} \sum_{j\neq k} \frac{w_{ij} + w_{ik}}{2 \overline{w_i}} a_{ij}a_{ik}a_{jk}
\end{split}
\end{equation}
thus defining the intensity of triangle $\Delta_{ijk}$ as $I^B_{\Delta ijk} = \frac{w_{ij} + w_{ik}}{2 \overline{w_i}} a_{ij}a_{ik}a_{jk}$ as the function of two of the triangle's weights and the average weight of the edges connected to node $i$, $\overline{w_i}$.

Proposed a bit later Onnela's definition~\cite{Onnela2005} scales the binary clustering by the average intensity of the triangles (see Appendix \ref{app:onnela} for more details):
\begin{equation}
	C^O_i = \frac{\left(W^{\left[\frac{1}{3}\right]}\right)^3_{ii}}{d_i (d_i - 1)} = C^{\text{bin}}_i \overline{I^O_{\Delta ijk}}
\end{equation}
with the triangle intensity defined as $I^O_{\Delta ijk} = \left( w_{ij} w_{ik} w_{jk} \right)^{1/3}$ and the average intensity taken over all triangles to which $i$ participates.

\fs{For all hybrid methods the denominator relies on the node's degree, meaning that the addition or deletion of edges will always significantly affect the clustering coefficient even if the edge has an infinitely small weight.}
Such methods can thus lead to inaccurate results when applied to the inferred networks, where a significant fraction of edges are false positives with small weights. 
In the following we will also demonstrate that they cannot reliably detect the most strongly clustered nodes in structured networks.
Only fully-weighted definitions can rise up to these challenges.

\textbf{Fully-weighted definitions}
are variants of the clustering coefficient that do not include any binary measures (anything that can be derived from the adjacency matrix alone, e.g. degrees).
In addition to substituting the number of triangles by the sum of triangle intensities, they also move away from counting triplets --- defining the maximum number of possible triangles $\max(n_{\Delta, i}) = n_{T,i} = d_i (d_i - 1)$.
Instead, they introduce the triplet intensity ($I_T$) such that, for a node $i$, $I_{T,i}$ is as a real-valued function of the weights associated to $i$.

One of the first fully-weighted definition for the clustering coefficient was provided by Zhang and Horvath \cite{Zhang2005} to analyze gene co-expression networks:
\begin{equation}
	C^Z_i = \frac{\sum_{j \neq k} w_{ij} w_{ik} w_{jk}}{\sum_{j \neq k} w_{ij} w_{ik}} = \frac{\overline{I^Z_{\Delta ijk}}}{\overline{I^Z_{T ijk}}} C^{\text{bin}}_i. \label{eq:Zhang_u}
\end{equation}
Note, that the fact that the definition can be expressed as a function of the binary clustering does not contradict the fully-weighted nature of the measure, as it stems from a simple recombination of the terms.

This definition can be interpreted as the ratio of the summed intensities $I^Z_{\Delta ijk} =w_{ij} w_{ik} w_{jk} $ of the triangles $\Delta_{ijk} = (i, j, k)$ to the maximal possible summed intensities $I^{Z(max)}_{\Delta ijk} = I^Z_{T ijk} = w_{ij} w_{ik}$ if all existing triplets $T_{ijk}$ were closed by an edge of weight 1 (the maximal possible weight in the normalized network).
This way, if $i$ is involved in a single triangle, the clustering coefficient is equal to the weight of the edge closing the triplet centered on $i$ (see also Table~\ref{tab:continuity}).

Though this definition does not fulfill the \emph{continuity} property, we will show that it still provides a consistent interpretation of weighted clustering, as discussed in~\cite{Kalna2006}, and is well suited to tackle networks with a large fraction of false positives. 

Other fully-weighted definitions that were proposed and discussed since \cite{Holme2007, Miyajima2014, Wang2016} do not bring significant additions compared to Zhang--Hovart's definition while actually losing some of its properties and its straightforward interpretation.
They are not considered further in this study --- see \suppmat{} \ref{sec:suppmat-weighted-old} for further explanations.

\begin{table}[]
	\centering
	\begin{ruledtabular}
		\begin{tabular}{r c c c c c c c c}
			&\includegraphics[width=2em]{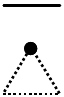}&\includegraphics[width=2em]{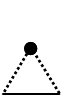}&\includegraphics[width=2em]{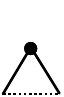}&\includegraphics[width=2em]{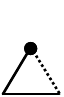}&\includegraphics[width=2em]{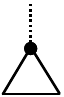}&\includegraphics[width=2em]{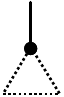}&\includegraphics[width=2em]{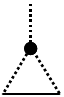}&\includegraphics[width=2em]{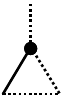}\\
			\hline
			$C^B$ & 1 & 1 & 1 & 1 & 1/2 & 0 & 1/3 & 1/2\\
			$C^O$ & 0 & 0 & 0 & 0 & 1/3 & 0 & 0 & 0\\
			$C^Z$ & 0 & 1 & 0 & 1 & 1 & 0 & 1/3 & 0\\
			$C$ & 0 & 0 & 0 & 0 & 1 & 0 & 0 & 0
		\end{tabular}
	\end{ruledtabular}
	
	\caption{Limit values for undirected weighted clustering coefficients of vertex $i$ (full circle) for different weight configurations in graphs with vanishing weights. 
		Solid lines depict edges of weight $w = \max(w) = 1$, dotted lines denote edges with vanishing weight $\epsilon$.
		Only the new continuous clustering (bottom row) returns the values consistent with the a continuity condition, whereas the definitions of Barrat ($C^B$),  Onnela ($C^O$), and Zhang--Horvath ($C^Z$) deviate for it.
	}
	\label{tab:continuity}
\end{table}

\subsection{A continuous definition for weighted clustering and closure}

For an undirected graph $G$, we define the new \emph{continuous clustering} of node $i$ as:
\begin{equation}
	C_i = \frac{\sum_{j \neq k} I_{\Delta i j k}^2}{\sum_{j\neq k} I_{T ijk}}
	    = \frac{\sum_{j \neq k} \left( \sqrt[3]{w_{ij} w_{jk} w_{ik}}\right)^2}{\sum_{j \neq k} \sqrt{w_{ji} w_{ik}}}. 
\end{equation}
We define the weighted intensity of triangles and triplets, respectively $I_{\Delta i j k} = \sqrt[3]{w_{ij} w_{jk} w_{ik}}$ and $I_{T ijk} = \sqrt{w_{ji} w_{ik}}$, using the geometric mean of the weights involved.
Thank to this, one strong weight in a triangle or triplet cannot compensate the presence of smaller weights, contrary to what may happen if one uses the arithmetic mean.
This provides the desired property that the intensity of triangles and triplets will go to zero if even a single edge weight goes to zero.
Note that, though the triangle intensity is defined as the geometric mean of the three weights involved, it is the square of this intensity that is used in the definition. 
The reason for this choice is twofold: to assign higher influence to large triangles and to ensure the linearity of the coefficient. 
Importantly, this definition of the triangle intensity $I_\Delta$ assigns the same role to all participating edges.
The new clustering could be interpreted as the ratio of the triangle intensity that is invested in strong triangles (given by the sum of the squared intensities of triangles, which increases the importance of strong triangles) to the triplet intensity, which would represent the maximum possible triangle intensity if all weights connecting adjacent nodes were equal to one.

\fs{The new continuous definition fulfils all the conditions we put forward above, and to he best of our knowledge it is the only one to do so.}
Similarly to previous definitions, the new clustering coefficient can also be rewritten in terms of node properties:

\begin{equation}
	C_i = \frac{\left( W^{\left[ \frac{2}{3} \right]} \right)_{ii}^3}{\left( s^{\left[ \frac{1}{2} \right]}_i \right)^2 - s_i}
\end{equation}
where $s_i = \sum_{j \neq i} w_{ij}$ is the normalized \textit{strength} of node $i$ and $W^{\left[ \alpha \right]} = { w_{ij}^\alpha }$ and $s^{\left[ \alpha \right]}_i = \sum_{j \neq i} w_{ij}^\alpha$, the fractional weight matrix and strength for any $\alpha \in \mathbb{R}$.

Same as for the previous definitions, the continuous clustering can be interpreted as a function of intensities and the binary clustering:
\begin{equation}
	C_i	    =	\frac{n_\Delta \overline{I^2_{\Delta ijk}}}{n_T \overline{I_{T ijk}}}
		  \;=\;	\frac{\overline{I^2_{\Delta ijk}}}{\overline{I_{T ijk}}} C^{\text{bin}}_i
		  \;=\;	\frac{\text{Var}(I_{\Delta ijk}) + \overline{I_{\Delta ijk}}^2}{\overline{I_{T ijk}}} C^{\text{bin}}_i,
\end{equation}
with means $\overline{I_{\Delta ijk}}$ and $\overline{I_{T ijk}}$ and  variance taken over all the triangles or triplets the node $i$ participate in respectively. 
In the limit where all triangles associated to node $i$ have similar intensities, we can neglect the variance term, leaving $\frac{\overline{I_{\Delta ijk}}^2}{\overline{I_{T ijk}}} C^{\text{bin}}_i$.
In this case, contrary to the Zhang--Horvath definition (Eq.~\ref{eq:Zhang_u}) the absolute value of the intensity matters, not only its ratio to the maximum possible intensity.
For a given average triangle intensity, the positive contribution of the variance implies that nodes with more variable intensities, i.e. at least one triangle with a high intensity, will have higher clustering coefficients than nodes with identical triangles of average-intensity.

Finally, the global clustering can also be defined in a straightforward fashion.
For simplicity, we define $\mathcal{I}_{\Delta, i} = \sum_{j\neq k} I_{\Delta ijk}^2$ and $\mathcal{I}_{T, i} = \sum_{j\neq k} I_{T ijk}$, leading to $C_i = \mathcal{I}_{\Delta, i} / \mathcal{I}_{T, i}$.
Using this definition, the continuous global clustering is obtained via the formula:
\begin{equation}
	C_g = \frac{\sum_i \mathcal{I}_{\Delta, i}}{\sum_i \mathcal{I}_{T, i}}
\end{equation}

\begin{table*}[t]
	\centering
	\begin{ruledtabular}
		\begin{tabular}{>{\centering\arraybackslash} m{0.09\textwidth} >{\centering\arraybackslash} m{0.23\textwidth} >{\centering\arraybackslash} m{0.12\textwidth} >{\centering\arraybackslash} m{0.15\textwidth} >{\centering\arraybackslash} m{0.18\textwidth} >{\centering\arraybackslash} m{0.23\textwidth}}
			\textbf{Mode} & \textbf{Pattern} & $\boldsymbol{n^{(m)}_{\Delta,i}}$ & $\boldsymbol{n^{(m)}_{T,i}}$ & $\boldsymbol{I_{\Delta, i}^{(m)}}$ & $\boldsymbol{I_{T, i}^{(m)}}$\\\hline
			Cycle & \raisebox{-.5\height}{\includegraphics[width=3.5cm]{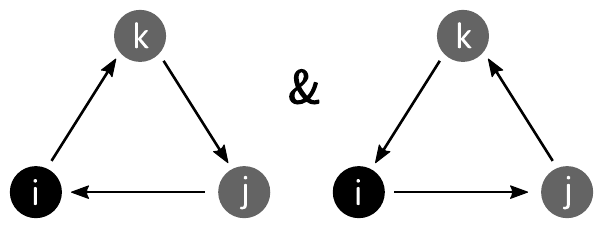}} & $A^3_{ii}$ & $\begin{array}{c}d_{i, in}d_{i, out} - d_i^\leftrightarrow \\ = \sum_{j \neq k} a_{ij} a_{ki} \end{array}$ & $\left(W^{\left[\frac{2}{3}\right]}\right)^3_{ii}$ & $s^{\left[\frac{1}{2}\right]}_{i, in}s^{\left[\frac{1}{2}\right]}_{i, out} - s^\leftrightarrow_i$\\
			Middleman & \raisebox{-.5\height}{\includegraphics[width=3.5cm]{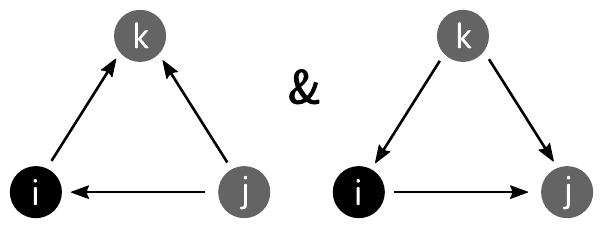}} & $\left( A A^T A \right)_{ii}$ & $\begin{array}{c}d_{i, in}d_{i, out} - d_i^\leftrightarrow \\ = \sum_{j \neq k} a_{ij} a_{ki} \end{array}$ & $\left(W^{\left[\frac{2}{3}\right]}W^{\left[\frac{2}{3}\right]T}W^{\left[\frac{2}{3}\right]}\right)_{ii}$ & $s^{\left[\frac{1}{2}\right]}_{i, in}s^{\left[\frac{1}{2}\right]}_{i, out} - s^\leftrightarrow_i$\\
			Fan-in & \raisebox{-.5\height}{\includegraphics[width=3.5cm]{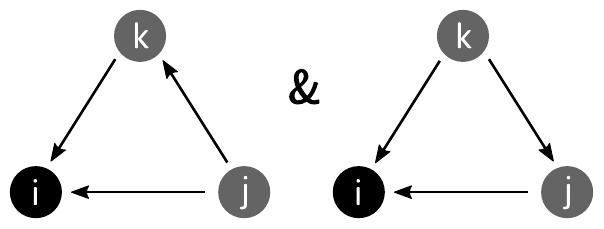}} & $\left( A^T A^2 \right)_{ii}$ & $\begin{array}{c} d_{i, in} (d_{i, in} - 1) \\ = \sum_{j \neq k} a_{ji} a_{ki} \end{array}$ & $\left(W^{\left[\frac{2}{3}\right]T}\left(W^{\left[\frac{2}{3}\right]}\right)^2\right)_{ii}$ & $\left(s^{\left[\frac{1}{2}\right]}_{i, in}\right)^2 - s_{i, in}$\\
			Fan-out & \raisebox{-.5\height}{\includegraphics[width=3.5cm]{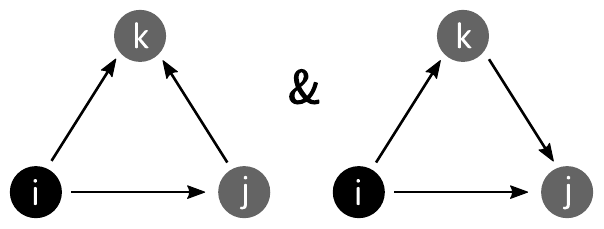}} & $\left( A^2 A^T \right)_{ii}$ & $\begin{array}{c} d_{i, out} (d_{i, out} - 1) \\ = \sum_{j \neq k} a_{ij} a_{ik} \end{array}$ & $\left(\left(W^{\left[\frac{2}{3}\right]}\right)^2 W^{\left[\frac{2}{3}\right]T}\right)_{ii}$ & $\left(s^{\left[\frac{1}{2}\right]}_{i, out}\right)^2 - s_{i, out}$
		\end{tabular}
	\end{ruledtabular}
	\caption{Definitions of the continuous intensities for each partial mode pattern in directed graph.
		Column 1: pattern names; column 2: patterns illustration; column 3: number of triangles for node $i$; column 4: number of triplets for node $i$; column 5: continuous intensities of the triangles for node $i$; column 6: continuous intensities of triplets for node $i$.
		The clustering coefficients associated to each mode $m$ are given by $C^{(m) \text{bin}}_i = n^{(m)}_{\Delta,i} / n^{(m)}_{T,i}$ for binary networks and $C_i^{(m)} = I_{\Delta, i}^{(m)}/I_{T, i}^{(m)}$ for the continuous definition.}
	\label{tab:partial-clustering}
\end{table*}

\subsection{Directed weighted clustering}

\fs{Fagiolo~\cite{Fagiolo2007} proposed how to generalize clustering to directed networks.}
He defined the different patterns or motifs (shown in Table~\ref{tab:partial-clustering}) that can exist in these networks, and adapted Onnela's definition~\cite{Onnela2005} as the first weighted directed clustering coefficient.
Barrat's definition~\cite{Barrat2004} was generalized to directed graphs in~\cite{Clemente2018} following the same distinction into Fagiolo's cycle, middleman, fan-in, and fan-out motifs.

\fs{Similarly, the Zhang--Horvath~\cite{Zhang2005} and continuous definitions can be generalized in a straightforward manner for directed networks.}
For it, we only need to redefine the intensities of each directed triangle and triplet motif as shown in Table~\ref{tab:partial-clustering} for the continuous definition and in Table~\ref{tab:partial-clustering_others} for Zhang--Horvath.
This simply requires replacing $A$ by $W$ in all expressions of $n^{(m)}_{\Delta,i}$ and $a$ by $w$ in all expressions of $n^{(m)}_{T,i}$.
As the total directed clustering is defined as the sum of all modes, we can write it as:
\begin{equation}
	C^{Z (tot)}_i = \frac{I^{Z (tot)}_{\Delta, i}}{I^{Z (tot)}_{T, i}} = \frac{(W + W^T)^3_{ii}}{\sum_{j \neq k} (w_{ij} + w_{ji})(w_{ik} + w_{ki})}
\end{equation}

Finally, the continuous clustering can be extended for each directed mode (see Table \ref{tab:partial-clustering}) and, for the total directed clustering, this leads to:
\begin{equation}
	C_i^{(tot)} = \frac{I_{\Delta, i}^{(tot)}}{I_{T, i}^{(tot)}} =  \frac{\frac{1}{2} \left(W^{\left[\frac{2}{3}\right]} + W^{\left[\frac{2}{3}\right]T} \right)^3_{ii}}{\left(s^{\left[\frac{1}{2}\right]}_{i, tot}\right)^2 - s_{i, tot} - 2 s^{\leftrightarrow}_i},
	\label{eq:c-dir}
\end{equation}
with $s_{i, tot} = \sum_{j} \left(w_{ij} + w_{ji} \right)$ the total strength, $s^{\left[\frac{1}{2}\right]}_{i, tot} = \sum_{j} \left( w_{ij}^{\frac{1}{2}} + w_{ji}^{\frac{1}{2}} \right)$ the total root strength, and $s^{\leftrightarrow}_i = \sum_j \sqrt{w_{ij} w_{ji}}$ the reciprocal strength.

As for the undirected case, the global clustering coefficient associated to each directed pattern can be obtained via the formula: $C_g^{(m)} = \sum_i I_{\Delta, i}^{(m)} / \sum_i I_{T, i}^{(m)}$.

\begin{figure*}[]
	\includegraphics[width=\textwidth]{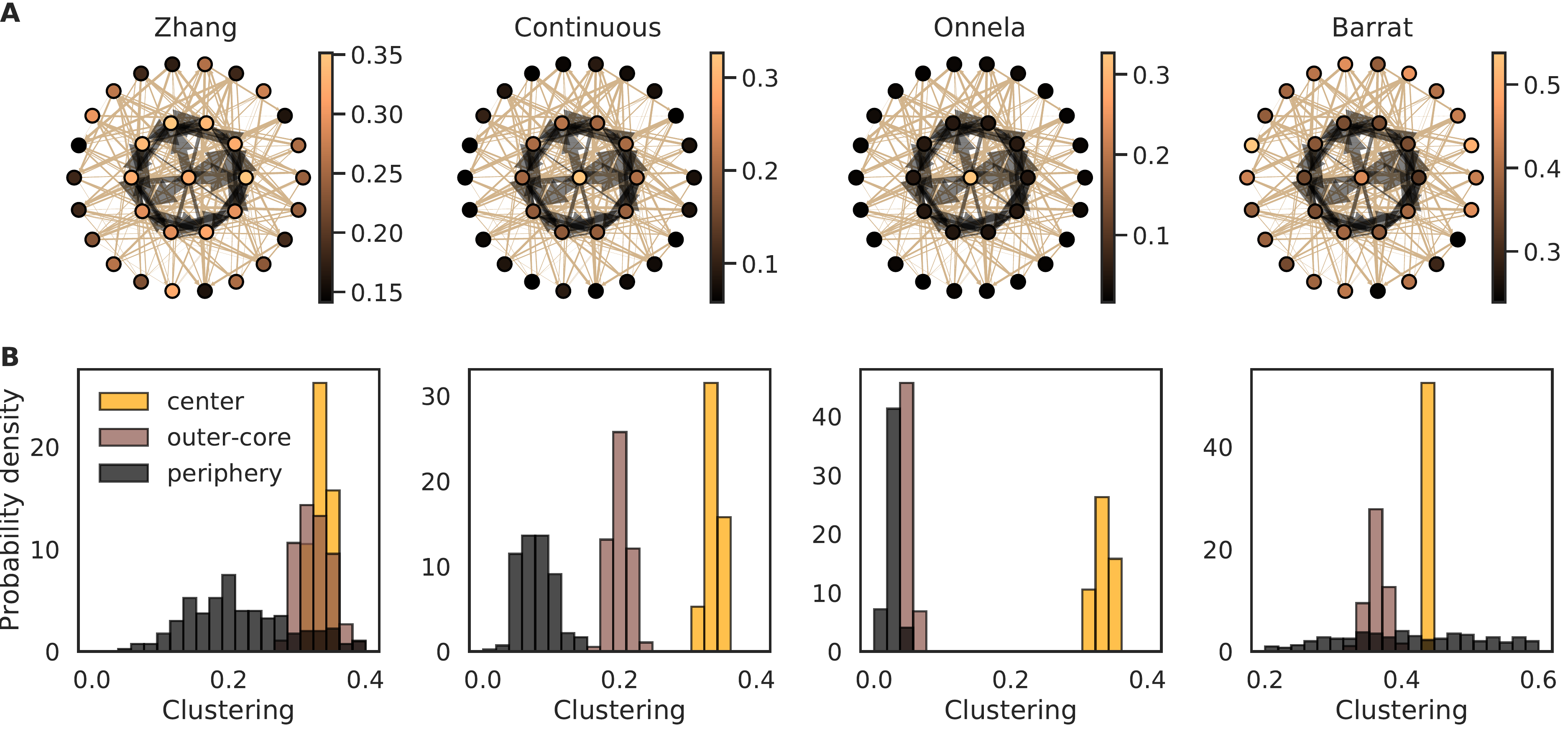}
	\caption{
		Only continuous clustering coefficient uncovers the true structure in the weighted core-periphery network. 
		A network has a 11 strongly-connected core nodes (black edges) that interact with well-clustered periphery nodes with weaker connection strengths (light-brown edges), see Appendix \ref{subsec:core-periphery} for details on the network.
		\textbf{A}. Graphical view of the network; edge width gives the strength of the connection, node color gives its clustering coefficient.
		\textbf{B}. Distribution of clustering coefficients for the three types of nodes over 10 realizations of such a core-periphery network.
		Only the continuous definition differentiates between the central, the outer-core and the periphery nodes.
		In all other methods, the clustering coefficients of the 10 ``outer-core'' and the 22 periphery nodes overlap: Onnela's definition only distinguishes the central node, while Barrat and Zhang--Horvath definitions do not hint at a core-periphery structure.
	}
	\label{fig:core-periph}
\end{figure*}

\section{The advantages of fully-weighted definitions}

\fs{In this section we discuss the sensitivity to the weight-encoded topological features and stability to noise in network measurements of the different clustering methods.}
A previous study~\cite{Saramaki2007} already noted the fact that previous definitions did not fulfill the \textit{continuity} condition by analyzing the behavior of the different coefficients for nodes that participate to a single-triangle.
Table~\ref{tab:continuity} illustrates some of these cases and shows that the new continuous definition is the only one to behave as expected.

Yet, we note that the definition of Zhang and Horvath is also very resilient to noise because, except for the corner-cases associated to single triangles, its behavior is continuous in all other situations.
Moreover, contrary to what was asserted in \cite{Saramaki2007}, it provides a perfectly sensible behavior given its interpretation of clustering as the ratio of the triangle intensity $I_{\Delta ijk}^Z = w_{ij} w_{ik} w_{jk}$ to its maximum possible intensity given the weights of node $i$: $I_{\Delta ijk}^{Z(max)} = I_{T ijk}^Z = w_{ij}w_{ik}$ if $w_{jk} = 1$. 

Because the definitions from Barrat's and Onnela's \cite{Barrat2004, Onnela2005} teams are the most well-known and (to the best of our knowledge) the only methods implemented in popular graph libraries, we restrict our comparison to Zhang--Horvath's and these two definitions.
A more comprehensive discussion of other definitions of the weighted clustering coefficient can be found in \suppmat{} \ref{sec:suppmat-weighted-old}.

\subsection{Sensitivity to weight-encoded topological features}

\begin{figure*}[]
	\includegraphics[width=\textwidth]{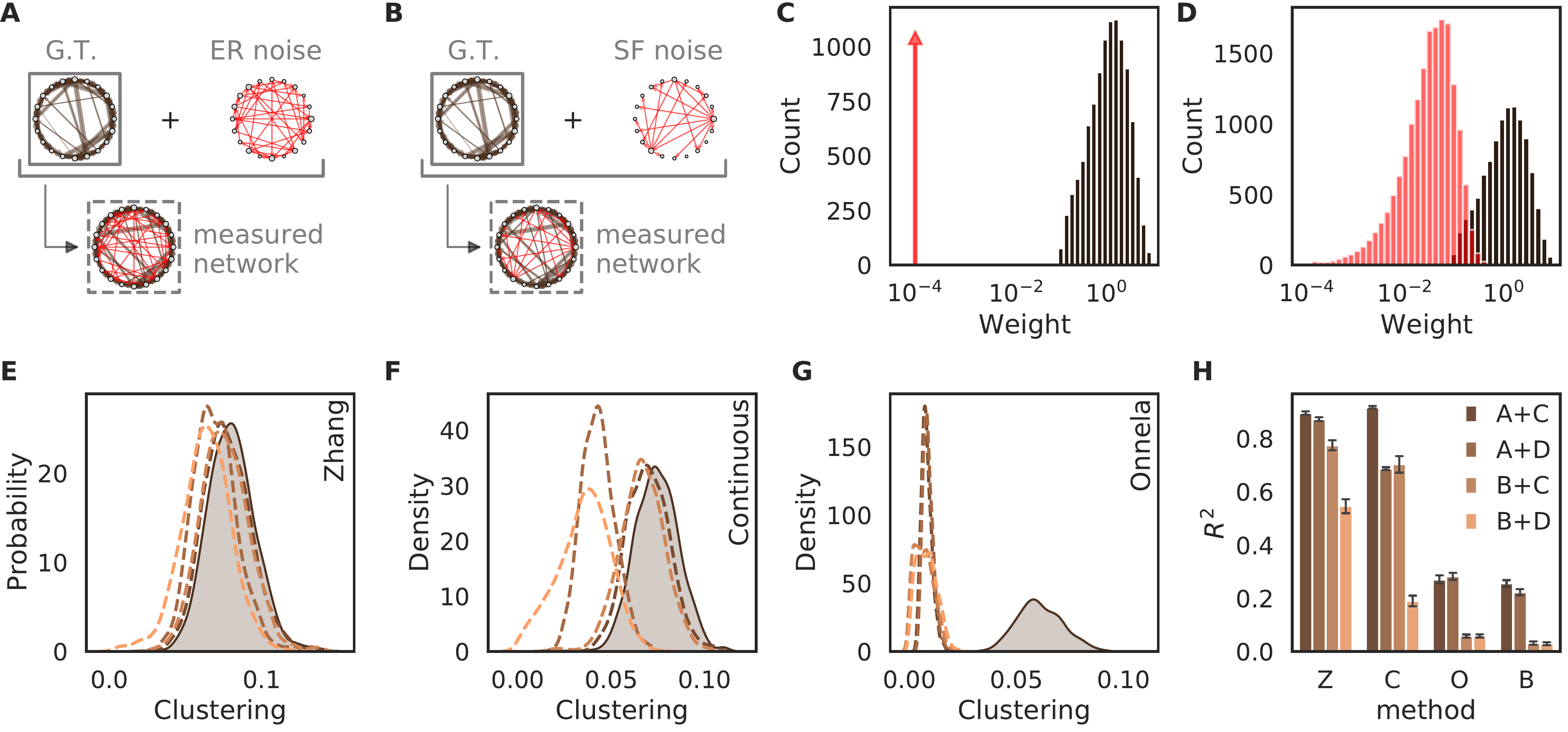}
	\caption{Fully-weighted methods are less sensitive to spurious edges.
		A ``measured network'' can be represented as the union of a ``ground truth'' (G.T.) --- here a Watts--Strogatz network in dark brown --- and spurious small-weight connections (``noise'' graphs with: \textbf{A} random, or \textbf{B} scale-free connectivity, in red).
		We assess the influence of the weight distribution of spurious connections (red) by checking weights that are: \textbf{C} all equal and small or \textbf{D} following an exponential distribution and overlapping with the real weights (dark brown).
		\textbf{E}--\textbf{G}: ground-truth clustering distribution (filled dark brown) compared to the distributions associated to the measured networks for each method (dashed lines). Weight and noise types, from (A+C) to (B+D), are associated to colors from brown to orange in the same order as in H.
		\textbf{H}: correlation between the ground truth clustering and clustering in measured networks for indicated spurious edge topology and weights.	
		Fully-weighted clusterings retain most of the correlation for (A+D), with $R^2 > 0.55$  and only lose the original information for (B+D).
		The results were obtained for 10 realizations of the spurious edges; error bars give confidence intervals.
	}
	\label{fig:spurious_edges}
\end{figure*}

\fs{Here we investigate how weighted structures can be detected ore missed using different clustering definitions.}
As an example we consider a weighted core-periphery graph, Figure~\ref{fig:core-periph}.
There, core nodes are characterized by both a dense binary connectivity and large weights, whereas periphery nodes display both sparser connectivity (though they still have large degrees) and weaker weights, for more details in the see Appendix~\ref{subsec:core-periphery}.
We generate 10 realizations of the network and consider distribution of clustering coefficients of different types of nodes. 
Continuous definition leads to distinct clustering of different type of nodes, making the true structure of the network clearly visible already in the clustering distributions. 
Because of their hybrid nature, Barrat's and Onnela's definitions cannot capture this underlying structure as it is mostly encoded in the weights and not at all in the degrees (core nodes are not binary hubs).
Though it is purely weighted, the Zhang--Horvath definition is also not suited to detect this type of weighted structure because its interpretation of a node's clustering only accounts for the relative triangle intensity given the node's weights.

\fs{The continuous clustering is sensitive to any topological property that is encoded via specific weight distributions.}
Furthermore, contrary to the Zhang--Horvath definition, it accounts not only for the ratio of the triangle intensity over the triplet intensity (how strong are the triangles compared the maximum possible value given the node's weights) but also for the absolute value of the intensity: a weak triangle, even if it corresponds to the highest possible value given the node's weights, will decrease the node's clustering in the continuous definition whereas it increases it with the Zhang--Horvath method.
In that sense, the continuous clustering provides a more global evaluation of the clustering coefficient compared to Zhang--Horvath that provides a more local information.

The definition from Barrat~\textit{et al.} has several limitations because it is close to being weight-insensitive \cite{Onnela2005, Antoniou2008}. 
It is particularly unsuitable for assessing networks with a potentially large number of low-weight spurious connections or very heterogeneous weight distributions. 
For this reason, we will mostly leave this clustering aside in the rest of this study.

\subsection{Continuity and resilience to noise}

\fs{Stability of a network measure to noise is of a particular importance for networks that are obtained via experimental measurements since these are often subject to noise and statistical biases, notably for inferred networks.} 
Methods abiding by the \textit{continuity} condition are especially resilient to the presence of low-weight spurious edges.
Violation of continuity can have significant and pervasive consequences for inference of network properties in many network structures.
We have already seen the simple examples in Table \ref{tab:continuity}, here we demonstrate that they are not just corner cases, but occur in larger, real-world networks. 

\fs{We illustrate the impact of spurious edges on measured clustering coefficients using the example of Watts-Strogatz small-world networks.} 
We consider different topologies for the subnetwork formed by the spurious edges: either an Erd\H{o}s-Renyi random network (Figure~\ref{fig:spurious_edges}~A.), associated to uncorrelated noise, or a scale-free network (Figure~\ref{fig:spurious_edges}~B.), which would correlate noise with certain nodes in the network.
Additionally, weights on the spurious edges could be much smaller than the weight of the actual edges (Figure~\ref{fig:spurious_edges}~C.) or have an overlapping distribution (Figure~\ref{fig:spurious_edges}~D.). 
Both fully-weighted methods are unaffected by low-noise (conditions A+C and B+C) and are also less influenced by the spurious edges when they weights are large enough to overlap with the real weight distribution (conditions A+D and B+D).
On other hand, because hybrid methods explicitly depend on the nodes' degrees, they are very susceptible to the presence of spurious edges  --- Figure~\ref{fig:spurious_edges} G, H.

\fs{The difference in behavior between the methods can be easily explained by a first-order expansion.} 
We consider change in the clustering coefficient of a node $i$ with degree $d_i$ after addition of a spurious edge $e = (i, v)$ with weight $\epsilon \ll 1$. 
For Barrat's and Onnela's methods, the new clustering coefficient becomes:
\begin{eqnarray*}
	C^{O'}_i &=& \frac{I_{\Delta, i}^C + O\left(\epsilon^{\frac{1}{3}} \right)}{d_i (d_i + 1)} = \frac{d_i - 1}{d_i + 1} C^O_i  + O\left(\epsilon^{\frac{1}{3}}\right) \overset{\epsilon\rightarrow 0}{\nrightarrow} C^O_i \\
	C^{B'}_i &=& \frac{I_{\Delta, i}^B + \sum_k \frac{\epsilon + w_{ik}}{2} a_{vk}a_{ik}}{I_{T, i}^B + s_i + \epsilon} =   \frac{d_i - 1}{d_i} C^B_i+ O(\epsilon)  \overset{\epsilon\rightarrow 0}{\nrightarrow} C^B_i
\end{eqnarray*}
meaning that, for both methods, the coefficients will deviate from the original clustering $C^{O/B}_i$ by a non-infinitesimal value, even when the perturbation was infinitesimal --- see \suppmat{} \ref{app:hybrid} for complete derivation.

On the other hand, the continuous clustering becomes
\begin{equation}
\begin{split}
C_i	&=\frac{I_{\Delta,i} + \sum\limits_{k\sim i} \left( \epsilon \, w_{vk} w_{ki} \right)^{\frac{2}{3}}}{I_{T,i} + 2s^{\left[\frac{1}{2}\right]}_i\sqrt{\epsilon}}\\
	&= C_i \left( 1 - \frac{2s^{\left[\frac{1}{2}\right]}_i\sqrt{\epsilon}}{I_{T,i}} + O\left( \epsilon^{\frac{2}{3}} \right) \right) \\
	&=  C_i + O\left( \sqrt{\epsilon} \right) \xrightarrow[\epsilon\rightarrow 0^+]{} C_{i}
\end{split}
\label{eq:continuous_continuity}
\end{equation}
showing only an infinitesimal deviation to the similarly infinitesimal perturbation.

Similarly, except for the single-triangle cases discussed in \ref{tab:continuity}, the Zhang--Horvath clustering becomes
\begin{equation}
	C^{Z'}_i = \frac{I_{\Delta,i}^Z + O(\epsilon)}{I_{T,i}^Z + O(\epsilon)} = C^Z_i  + O(\epsilon)  \xrightarrow[\epsilon\rightarrow 0^+]{} C^Z_i
\end{equation}

It is worth noting that one \emph{continuity} issue, associated to nodes participating in only one triangle, occurs for all definitions but the continuous one.
Since this situation is pervasive in networks with low degree or binary clustering, using the continuous clustering definition can be of particular importance in such cases --- see Appendix~\ref{app:single-tr}.

\begin{figure*}[t]
	\includegraphics[width=\textwidth]{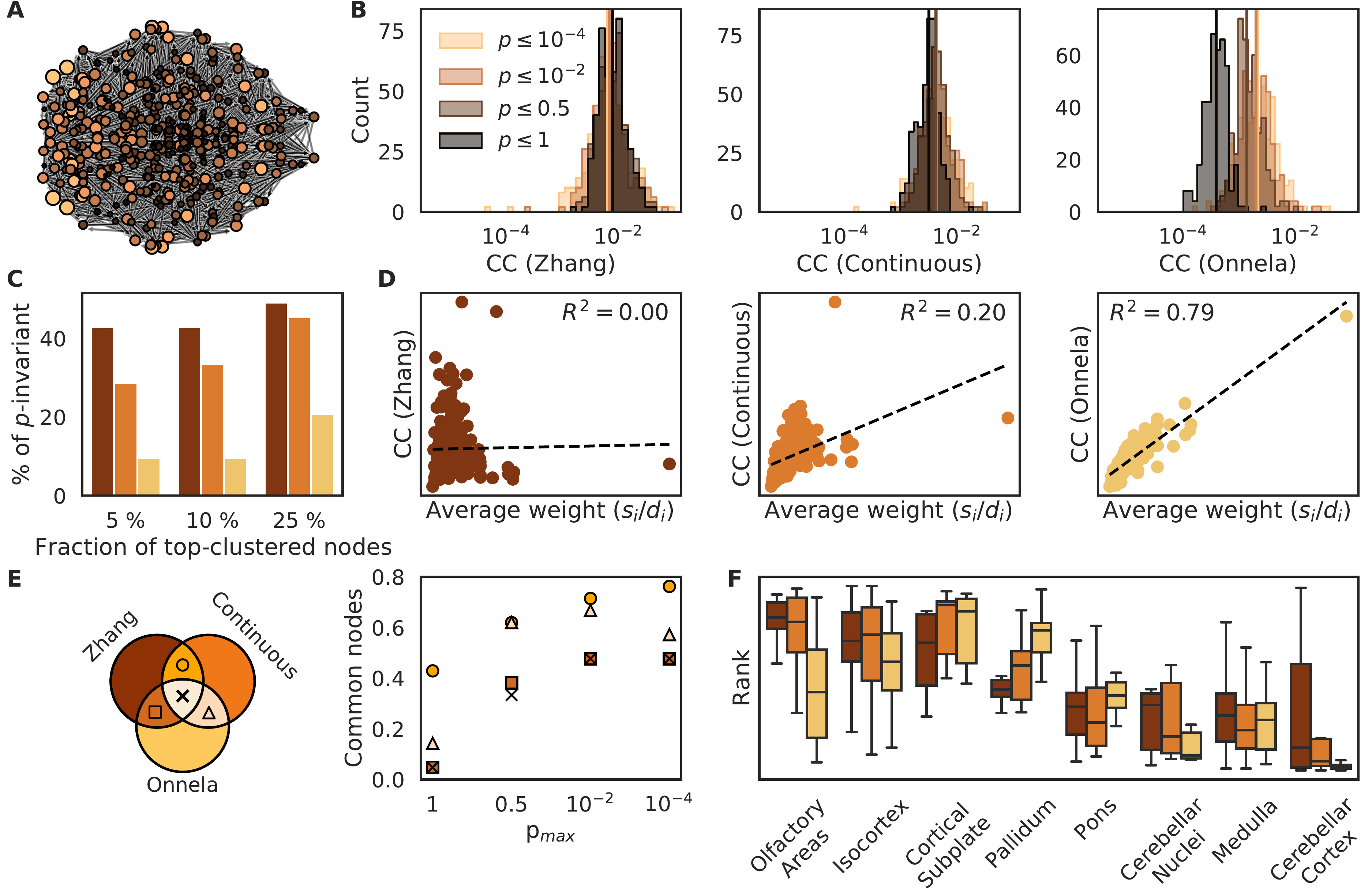}
	\caption{
		For the mouse connectome different clustering methods give significantly different results. 
		\textbf{A}. Top view of the mouse brain, node size -- total-degree,  node color --  out-degree (lighter colors for higher degrees).
		\textbf{B}. Distribution of the total clustering coefficients if only edges with the p-value $p < p_\mathrm{max}$ are preserved. Smaller changes in clustering distribution for fully-weighted definitions than for Onnela.
		From yellow to black thresholding keeps respectively 9, 13, 32, and 100\% of the original network.
		\textbf{C}. The fraction of the nodes with highest total clustering (top 5, 10 and 25\%) that are preserved across all subsamplings in B for Zhang (brown), continuous (orange), and Onnela (pale yellow).
		\textbf{D}. Correlation of the three total clustering definitions with average total node weight ($s_i/d_i$) shows that fully-weighted definitions captures additional information beyond degree and strength.
		\textbf{E}. Fraction of the 10\% highest clustering nodes that are common between two of the definitions (full markers on right panel include the central region of the left panel) or among all three definitions (black crosses, central region) as shown on the Venn diagram.
		\textbf{F}. Clustering ranks of the areas within brain regions (telling which regions contain nodes with high clustering coefficients) can significantly vary depending on the definition (Zhang - brown, continuous - orange, and Onnela - pale yellow).
	}
	\label{fig:mouse}
\end{figure*}

\begin{figure*}[t]
	\includegraphics[width=\textwidth]{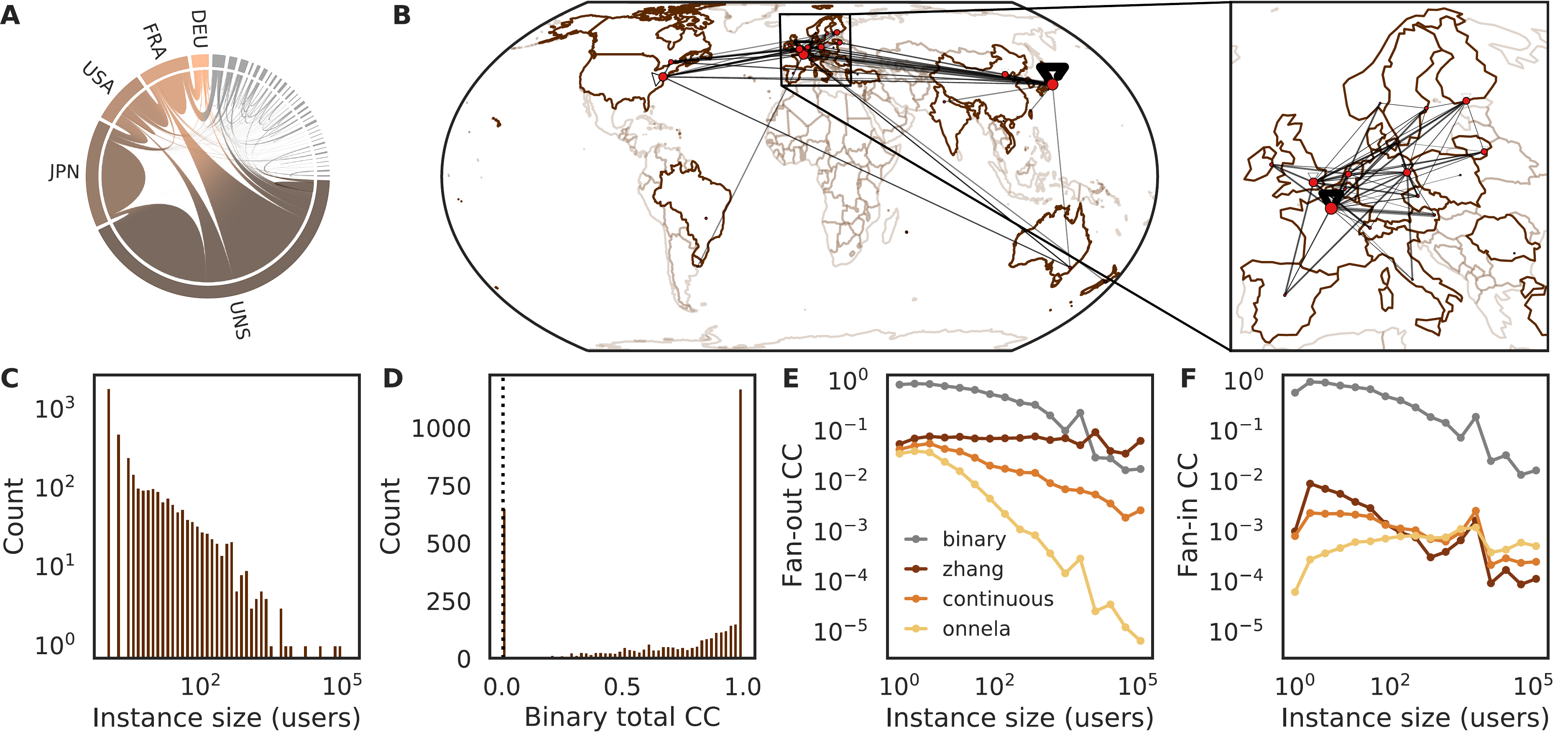}
	\caption{
		Properties of the network of Fediverse instances.
		\textbf{A}. Chord diagram of connections between locations.
		\textbf{B}. Spatial representations of the network showing connections that amount for at least 1\textperthousand{} of the strongest connection,  nodes placed on each country's capital, their sizes represent the number of instances hosted in that country.
		The zoom on Europe show all connections in the European subnetwork.
		\textbf{C}. The counts of users per instance follows a heavy-tailed distribution.
		\textbf{D}. The network displays strong structural clustering, most nodes with non-zero in- and out-degrees displaying binary clustering values close two one, whereas the expected value for an Erd\H{o}s-Renyi network with the same number of edges would be almost zero (dotted line).
		\textbf{E}. The median values of the fan-out clustering for different instance sizes show that the strong heterogeneity of the network can have a notable influence for the Onnela method (yellow) whereas Zhang--Horvath (dark brown) and continuous methods (orange) display much weaker correlation with the size of the instance.
		\textbf{F}. The fan-in motifs have intensities that are several orders of magnitude lower than fan-out and display weaker dependency on the instance size. The binary (gray) clustering coefficient misses the difference between fan-out and fan-in, which predominantly relies on the weights' effect. 
	}
	\label{fig:fediverse}
\end{figure*}

\begin{figure*}[t]
	\includegraphics[width=\textwidth]{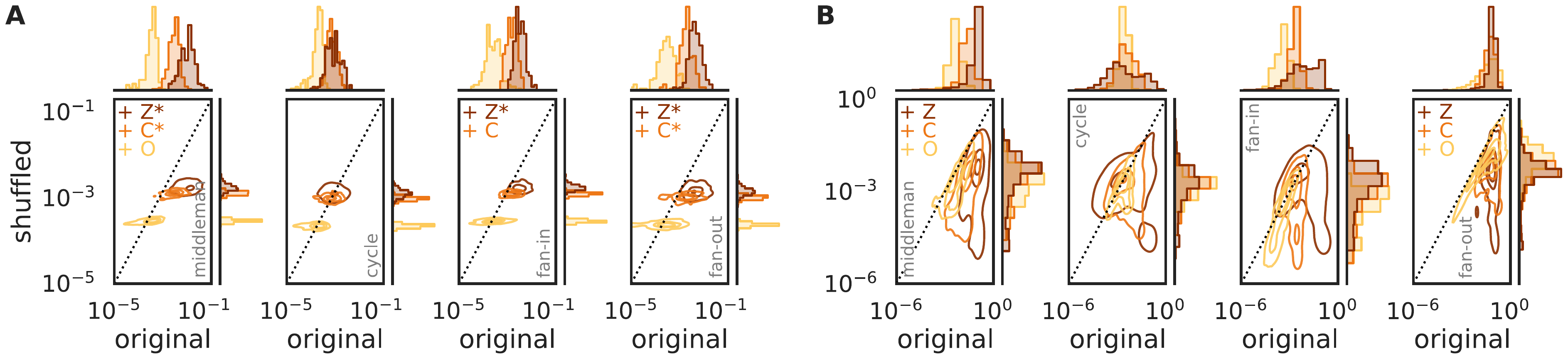}
	\caption{
		Different structures of clustering patterns and information flow in the mouse brain and the Fediverse.
		The original clustering values are compared to graphs with the same adjacency matrix but shuffled edge weights (mouse, 10 uniformly shuffled networks; Fediverse, 200 graphs shuffled across all out-going edges of each node to preserve out-strength normalization).
		The contours show different density levels of the point clouds associated to the original/shuffled pairs of clustering coefficients.
		\textbf{A}. The Zhang--Horvath and continuous definitions capture the redundancy of information flow in the mouse brain (middleman and fan-in/fan-out motifs are higher in the original graph).
		However, the Onnela method does not capture this feature.
		\textbf{B}. For the Fediverse, only fan-out and middleman motifs are significantly stronger than in the random graphs.
		Patterns where the original values are significantly greater (resp. smaller) than the randomized ones are marked by the $+$ (resp. $-$) and the initial of the method (brown, Z for Zhang; orange, C for continuous; yellow, O for Onnela).
		The original clustering is considered to be significantly higher (resp. lower) if 75\% of the points where above (resp. below) the dotted identity line.
		Additional * denotes one-sided fractions greater than 95\%.}
\label{fig:kde-clst}
\end{figure*}

\section{Application to real world networks}

\subsection{Mouse mesoscale connectome}

\fs{In neuroscience, the networks on different scales give a vital piece of information to understand the brain better.} 
Unsurprisingly, connectomics, the mapping of the connections in the nervous system, gained significant attention and developed dramatically over the last years. 
Most of the networks in neuroscience are weighted, with all obtained connectivities are either measured or inferred, making them a typical example for the challenges discussed above.
The mouse mesoscale connectome~\cite{Oh2014} is a fascinating example of such networks, both because it provides information about the entire mouse brain and because it contains an evaluation of the probability of false positives for all connections.

\fs{Here, we investigate how the choice of the clustering coefficient definition can alter the results.}
The network is very inhomogeneous, with broadly varying degrees, Figure~\ref{fig:mouse}~A. 
The edges in the mesoscale connectome are assigned p-values that quantify their probability to correspond to real physical connection ($p$ denotes the probability of the connection to be a spurious edge).
They have therefore different significance levels, with only 13\% of all edges having $p$-values smaller than $0.01$, i.e. only 13\% of all edges have a probability to be spurious that is lower than 1\%.
We consider a thresholding procedure, where at each level of the threshold ($p_\mathrm{max}$), only the edges with smaller $p$-values are kept. 
This procedure can be seen as an attempt to remove spurious edges, though no correct threshold level is known.
Compared to the hybrid method, both fully-weighted definitions are much less sensitive to thresholding, Figure~\ref{fig:mouse}~B,~C.
Thus we see that the resilience to noise we showed analytically and on toy networks is relevant for a real-world network.
Furthermore, the resilience is not limited to the general shape of the distribution but indeed preserves the precise values and ranks: a larger fraction of the nodes displaying high clustering in the full graph are still among the highest ranking nodes in the thresholded graphs when fully-weighted method's are used compared to the hybrid method, Figure~\ref{fig:mouse} E.

\fs{Clustering coefficient is a network measure that captures features beyond purely local parameters, such as degree or strength}.
However, as we see for the mouse connectome, the hybrid method strongly correlates with the average weight associated with a node $i$: $s_i / d_i$.
At the same time, the fully-weighted definitions are much less correlated with it, Figure~\ref{fig:mouse} D. 
As a result, they can bring more independent information regarding weighted network structure than the hybrid method.
This trend is even stronger for the Zhang--Horvath definition, which does not account for the absolute intensity of triangles. 
In contrast, the continuous definition provides some intermediate behavior as the intensity of triangles often correlates, if only in part, with the average weight associated with the node.

Finally, though the continuous and Zhang--Horvath definition often provide somewhat similar results, they may  differ significantly, e.g. for cerebellar cortex on Figure~\ref{fig:mouse} F.
Combining the results of both methods can thus be informative, for example to single out nodes that possess only weak connections (and will therefore register as weakly clustered for the continuous definition) yet connect to other nodes that are strongly connected (thus registering as strongly clustered for Zhang--Horvath).

\subsection{Decentralized social media: the Fediverse}

The \href{https://fediverse.party/}{Fediverse} is a set of federated social media that can communicate via a collection of \href{https://en.wikipedia.org/wiki/Fediverse#Communication_protocols_used_in_the_fediverse}{common protocols}, the most well-known being ActivityPub.
This network can be seen as a set of alternatives to corporate platforms such as Facebook or Twitter.
Social media on the Fediverse usually promote ideas of decentralization, interoperability, free/libre and open-source software (FLOSS), and the absence of algorithmic filters in favor of human curation and moderation.

We analyze here a snapshot of this network that was obtained in 2018 by Zignani \textit{et al}. \cite{Zignani2019}\footnote{data available at \href{https://dataverse.mpi-sws.org/dataset.xhtml?persistentId=doi:10.5072/FK2/AMYZGS}{https://dataverse.mpi-sws.org/dataset.xhtml?persistentId=doi:10.5072/FK2/AMYZGS}}.
Contrary to the original publication, we chose here to look at the mesoscale level, i.e. at connections between instances (the equivalent of a community on the Fediverse, where at least one, but up to several thousand users can have an account).
This mesoscale view leads to a network of weighted directed interactions between communities of strongly connected users. 
Indeed, users of a single instance (the technical name for a community on the Fediverse) can see and interact with all public messages posted by other members on that same server. 
At the same time, they can only see a subset of the posts from people on other servers (either because they follow their author or because other members of the instance follow the author or shared this specific post).

For each instance $I_1$, an edge towards another instance $I_2$ means that at least one user on $I_1$ follows at least one member of $I_2$.
The precise value of the weight associated to this edge gives the fraction of all followers from the source instance that are associated to members of the target instance.
The weights are thus a proxy to characterize the fraction of the community's attention that is associated to content produced by another community; this means of course that, for each node, it's outgoing strength $s_o$ is equal to one.
Note that in this network, information flow occurs in the direction \emph{opposite} to that of the edge, because the directed edge denotes that the source is paying attention to what the target posts.

The network snapshot contains 3,825 nodes corresponding mostly to instances running \href{https://joinmastodon.org/}{Mastodon}, one of the most prominent microblogging platforms on the Fediverse.
These 3,825 nodes represent more than half of the entire network and are connected via 81,371 edges.
Chord diagram of connections between locations shows that, besides the fact that the position of most instances is unspecified (UNS), the largest hosting countries are Japan, the USA, and France, with Japanese and French communities interacting mostly among themselves, Figure~\ref{fig:fediverse} A, B.
The network is both very sparse and strongly heterogeneous, with a median degree of 5 but node degrees and sizes varying over 3 to 5 orders of magnitude.
This broad distribution has notable implications for different clustering definitions.
For the Zhang--Horvath, it increases the likelihood of running into the corner-cases of single triangles, increasing the average clustering compared to the other methods.
For Onnela's definition, it strongly correlates clustering values to the degree (and thus to the size) of the instance.

As for the mouse network, all weighted methods lead to results that differ strongly from the binary clustering, Figure~\ref{fig:fediverse} (D and E).
Some of the results from the hybrid method tends to correlate strongly with some 1st order properties of the nodes (Figure~\ref{fig:fediverse}~E) while the fully-weighted methods bring more independent information.
The Fediverse network displays a peculiar feature as its fan-out and fan-in clustering differ significantly despite the usual correlation between these two patterns, as can be seen from the comparison of Figure~\ref{fig:fediverse} E and F.

\subsection{Using local clustering to infer dynamical properties}

\fs{Analysis of clustering coefficient for different structured patterns offer a way to obtain a precise idea of the critical dynamical patterns within a network.}
\fs{To determine the significance of a particular pattern we compare its prominence in the original network with it in a null-model networks obtained via appropriate  randomization.}
For the mouse brain, as a randomized control, we take the original network and only shuffle the weights, thus preserving the weights distribution and binary structure.
Comparing the actual values in the original graphs to that of randomized graphs, we can see the importance of looking at fully-weighted measures.

\fs{Both Zhang--Horvath and continuous definitions identify the preference of the mouse brain network for redundant information flows, whereas the hybrid method of Onnela does not capture this feature.}
Redundant information transfer in the brain is indeed associated to situations where a signal can be transferred not only directly from one node to another, but also indirectly via a third (middleman) node, as can be seen in middleman, fan-in and fan-out patters (cf. Table~\ref{tab:partial-clustering}).
This overexpression of redudant patterns is visible, for the fully-weighted methods, in the high values of middleman and fan-in/fan-out motifs in Figure~\ref{fig:kde-clst}~A.

The situation for the Fediverse is significantly more complex.
Indeed, 98 \% of all edges belong to at least one triangle, some are involved in up to thousands of triangles.
A deeper understanding of the clustering requires precisely investigation of how the weight distribution correlates with specific patterns of triangles.
For instance, though fan-in and fan-out patterns co-occur and are thus usually correlated, the Fediverse displays an unexpected discrepancy between the weight associated to both patterns, as seen on Figure~\ref{fig:kde-clst}~B, which notably differentiates it from the mouse connectome (see also \suppmat{}~\ref{app:closure-shuffled}).

\section{Discussion}

\fs{In this work, we introduced new directed weighted definitions for clustering analysis.}
Using analytic derivations, generated networks models and real data, we showed that the behavior of fully-weighted measures displayed enhanced sensitivity, selectivity, and robustness compared to hybrid measures.
To facilitate access to these measures, all clustering methods were implemented in Python and made compatible with the three main libraries in this language (\texttt{networkx}, \texttt{igraph}, and \texttt{graph-tool}) \cite{NNGT}.

\fs{We highlighted the importance of continuity as a crucial notion for networks with highly heterogeneous weight distribution or numerous spurious edges with low weights.}
In the previous studies, slight variations on this notion had been introduced mathematically \cite{Wang2016} and hinted at by the study of corner cases \cite{Saramaki2007}.
However, the continuity property was inadvertently considered only in particular classes of networks~\cite{Wang2016} (disregarding, for example, small degree cases), erroneously marking previous methods as continuous. 
The other study~\cite{Saramaki2007} asserted discontinuity of previous measures simply as a feature, without discussing its implications or ways to avoid it.
Our new proposal of fully continuous clustering methods based on simple mathematical principles and requirements is, therefore, the first to fully solve the issue of continuity.
In addition, we show in \suppmat{} \ref{app:closure} that these principles can be extended to other measures such as the local closure.

\fs{We discussed how each weighted definition is associated to a specific interpretation of weighted clustering as a function of the binary clustering, triangle, and triplet intensities (see Appendix~\ref{app:compare-prop} and Table \ref{tab:interpretations} for a summary).}
In turns, these interpretations are associated to specific properties for each clustering coefficient.
Combining mathematical analysis and concrete examples, we asserted that fully-weighted measures out-perform hybrid ones to evaluate clustering in networks with large numbers of spurious edges with small weights.
We expanded the results from previous studies comparing existing weighted definitions \cite{Kalna2006, Saramaki2007, Antoniou2008, Wang2016}, extending their definition to directed networks and providing complete mathematical justifications to previous observations regarding linearity and continuity.

Our analyses show that either of the fully weighted definitions may be preferred depending on the network properties.
For networks with large degrees, the Zhang--Horvath definition may be preferred if the number of low-weight spurious edges is very high as it is least susceptible to noise.
In networks with heterogeneous weight distributions where the absolute value of triangle strength is of interest, we showed that the continuous definition provides more relevant results than that of Zhang and Horvath.
Indeed, in networks involving fluxes of good or matter as well as for information processing (e.g. in brain or telecommunication networks), one may be interested in the absolute amount that can be transferred between nodes, making nodes with small weights of little relevance.
A similar issue occurs for networks where many nodes participate to a single triangle (see Appendix~\ref{app:single-tr}) making the Zhang--Horvath clustering non-continuous.
In such networks, the use of the Zhang--Horvath definition is likely to assign high clustering to single-triangle nodes with low-weights, whereas the continuous definition will not, as was shown on Figure \ref{fig:core-periph}.

Finally, we illustrate the usefulness of weighted clustering methods to investigate clustering on the example of a connectome and a decentralized social network.
The fully-weighted methods were especially suitable to reveal key differences in their weighted structural properties.
Indeed, we showed that middleman, fan-in, and fan-out patterns, characteristic of pathways enabling redundant information transfer between nodes, were overexpressed in the mouse brain.
In the Fediverse, the methods revealed an unexpected discrepancy between fan-in and fan-out modes, probably associated with social interaction patterns that would mandate further investigation.

\subsection*{Acknowledgments}

The research was funded by a Humboldt Research Fellowship
for Postdoctoral Researchers and a Sofja Kovalevskaja Award
from the Alexander von Humboldt Foundation, endowed by the
Federal Ministry of Education and Research.

\nocite{*}
\bibliography{references}

\appendix

\section{Limitations of other fully-weighted definitions}
\label{sec:suppmat-weighted-old}

See Table \ref{tab:compare-weighted} for a complete comparison of the fully-weighted clustering definitions.

\subsection{Holme \textit{et al}. (2007)}

Most studies consider the definition of Holme \textit{et al.} \cite{Holme2007} to be:
\begin{equation}
	C_i^H = \frac{\sum_{jk} w_{ij} w_{ik} w_{jk}}{\max_{ij}(w_{ij})\sum_{jk} w_{ij} w_{jk}}
	\label{eq:holme-wrong}
\end{equation}
which would make it inconsistent with the binary definition.
However the discussion in their paper states that \emph{consistency} was one of their requirements, letting us think that they actually meant to define it as
\begin{equation}
	C_i^H = \frac{\sum_{j\neq k} w_{ij} w_{ik} w_{jk}}{\max_{ij}(w_{ij})\sum_{j\neq k} w_{ij} w_{jk}}
\end{equation}
which would make it equal to the definition from Zhang and Horvath \cite{Zhang2005}, we will therefore not consider \ref{eq:holme-wrong} here.

\subsection{Miyajima and Sakuragawa (2014)}

The authors define a multitude of generalized clustering coefficients based on the use of an arbitrary function $h: \mathbb{R}^2 \rightarrow \mathbb{R}$ such that:
\begin{equation}
	C_i^{M,h} = \frac{\sum_{jk} h( h(w_{ij}, w_{ik}), w_{jk})}{\sum_{j\neq k} h(h(w_{ij}, w_{jk}), \max_{lm}(w_{lm}))}
\end{equation}
More specifically, Wang \textit{et al.} \cite{Wang2016} argue in favor of the use of a specific version using the harmonic mean:
\begin{equation}
	C_i^{M, hm} = \frac{\sum_{jk} \frac{2}{\frac{2}{\frac{1}{w_{ij}} + \frac{1}{w_{ik}}} + \frac{1}{w_{jk}}}}{\sum_{j \neq k} \frac{2}{\frac{2}{\frac{1}{w_{ij}} + \frac{1}{w_{ik}}} + \frac{1}{\max_{lm}(w_{lm})}}}
\end{equation}
However, this definition suffers from three major shortcomings:
\begin{itemize}
	\item despite what is asserted by the authors, it does not fulfill the \emph{continuity} condition;
	\item it is not locally linear, meaning that two nodes that have the same neighborhood but with all weights differing by a factor $\lambda$ will not have the ratio of their clustering coefficients equal to $\lambda$;
	\item it introduces an undesired asymmetry in the definition of the triangle intensity: for a given triangle $\Delta_{ijk}$, the computed intensity will be different for each node as it depends on which one is considered as $i$.
\end{itemize}

\begin{table}[h]
	\centering
	\begin{ruledtabular}
	\begin{tabular}{r c c c c c c c c}
	&\includegraphics[width=2em]{edge_case_1}&\includegraphics[width=2em]{edge_case_2}&\includegraphics[width=2em]{edge_case_3}&\includegraphics[width=2em]{edge_case_4}&\includegraphics[width=2em]{edge_case_5}&\includegraphics[width=2em]{edge_case_6}&\includegraphics[width=2em]{edge_case_7}&\includegraphics[width=2em]{edge_case_8}\\
	\hline
	$C^Z$ & 0 & 1 & 0 & 1 & 1 & 0 & 1/3 & 0\\
	$C^H$ & 0 & 1 & 0 & 1 & 1 & 0 & 1/3 & 0\\
	$C^{M,hm}$ & 0 & 1 & 0 & 1 & 1 & 0 & 1/3 & 0\\
	$C$ & 0 & 0 & 0 & 0 & 1 & 0 & 0 & 0
	\end{tabular}
	\end{ruledtabular}

\caption{Undirected weighted clustering coefficients of vertex $i$ (full circle) for different weight configurations. Solid lines depict edges of weight $w = \max(w) = 1$, whereas dotted lines are associated to edges with vanishing weight $\epsilon$. Only the new continuous clustering (bottom row) displays the required properties, compared to the definitions of Zhang ($C^Z$), Holme ($C^H$), and Miyajima ($C^{M, hm}$).}
\label{tab:compare-weighted}
\end{table}

\section{Comparison of clustering properties}
\label{app:compare-prop}

Multiple properties of the main definitions for the clustering coefficient are listed in Table \ref{tab:compare-all}.
These properties depend on the definitions of triangle and triplet intensity that are recapitulated in Table \ref{tab:intensities}.

\begin{table*}[]
	\centering
	\begin{ruledtabular}
	\begin{tabular}{r c c c c c}
	\textbf{Property} & Barrat & Onnela & Miyajima & Zhang & Continuous\\
	\hline
	Consistent with binary definition & X & X & X & X & X\\
	Normalized ($C \in [0, 1]$) & X & X & X & X & X\\\
	All weights participate to $I_\Delta$ & & X & X & X & X\\\
	All weights participate equally to $I_\Delta$ & & X & & X & X\\\
	Linear against local scaling of the weights & & X & & X & X\\
	Sensitive to weight permutations & X & & X & X & X\\
	For a triangle $\Delta = (i, j, k)$, $I_\Delta = I_T f(w_{jk})$ & & & & X\\
	Continuous & & & & & X
	\end{tabular}
	\end{ruledtabular}

\caption{Comparison of the different properties of the different clustering coefficients. Zhang--Horvath and the continuous definitions fulfill the maximum number of desirable properties. An X means that the method possesses this property.}
\label{tab:compare-all}
\end{table*}

\begin{table}[H]
	\centering
	\begin{ruledtabular}
	\begin{tabular}{r c c c c c}
	\textbf{Definition} & Triangle ($I_{\Delta ijk}$) & Triplet ($I_{T ijk}$)\\
	\hline
	Barrat & $\displaystyle\frac{w_{ij} + w_{ik}}{2 \overline{w_i}} a_{ij}a_{ik}a_{jk}$ & $d_i (d_i - 1)$ \\\\
	Onnela & $\left( w_{ij} w_{ik} w_{jk} \right)^{1/3}$ & $d_i (d_i - 1)$ \\\\
	Miyajima (hm) & $\displaystyle\frac{2}{\frac{2}{\frac{1}{w_{ij}} + \frac{1}{w_{ik}}} + \frac{1}{w_{jk}}}$ & $\displaystyle\frac{2}{\frac{1}{w_{ij}} + \frac{1}{w_{ik}}}$\\\\
	Zhang & $w_{ij} w_{ik} w_{jk}$ & $w_{ij} w_{ik} $\\\\
	Continuous & $\sqrt[3]{w_{ij} w_{jk} w_{ik}}$ & $\sqrt{w_{ij} w_{ik}}$
	\end{tabular}
	\end{ruledtabular}

\caption{Comparison of the formula for triangle and triplet intensity among the different clustering definitions for undirected networks.}
\label{tab:intensities}
\end{table}

The different formulas for the intensities lead to different interpretations of weighted clustering --- Table \ref{tab:interpretations}.
Barrat and Zhang only quantify ratios of triangle and triplet strength. while Onnela and the continuous definitions are sensitive to the absolute value of the triangle intensity. 
The latter provides an intermediate between Zhang and Onnela as it reacts to both the ratio of intensities and the absolute value of the triangle intensity.

\begin{table}[H]
	\centering
	\begin{ruledtabular}
	\begin{tabular}{r c c c c c}
	\textbf{Definition} & Barrat & Onnela & Zhang & Continuous\\
	\textbf{Formula} & $\displaystyle\frac{\overline{w^\Delta_i}}{\overline{w_i}} C^{\text{bin}}_i$ & $\displaystyle\overline{I^O_{\Delta ijk}} C^{\text{bin}}_i$ & $\displaystyle\frac{\overline{I^Z_{\Delta ijk}}}{\overline{I^Z_{T ijk}}} C^{\text{bin}}_i$ & $\displaystyle\frac{\overline{I^2_{\Delta ijk}}}{\overline{I_{T ijk}}} C^{\text{bin}}_i$
	\end{tabular}
	\end{ruledtabular}
\caption{Comparison of the interpretation of the different clustering definitions for undirected networks.}
\label{tab:interpretations}
\end{table}

\section{Derivation of the evolution of hybrid clustering coefficients}

\label{app:hybrid}

\subsection{Barrat}
\label{app:barrat}

Upon addition to a graph $G(N, E)$ of an edge $(i, v)$ of weight $\epsilon$, giving $G'(N, E' = E + \{(i, v)\}$, one can compute the evolution of the initial clustering coefficient $C^B_{i}$ to its new value $C^{B'}_i$ by rewriting the definition from \cite{Barrat2004} as in \cite{Saramaki2007}:

\begin{equation}
\begin{split}
	C^B_i	& = \frac{1}{d_i(d_i-1)} \sum_{j\neq k} \frac{w_{ij} + w_{ik}}{2 \overline{w_i}} a_{ij}a_{ik}a_{jk}\\
			& = \frac{1}{d_i(d_i-1)} \sum_{j\neq k} \frac{w_{ij}}{\overline{w_i}} a_{ij}a_{ik}a_{jk}
\end{split}
\end{equation}

From this, we can deduce the following relationship between Barrat's definition and the binary clustering coefficient $C^{bin}_i$:

\begin{equation}
	C^B_i = \frac{n_{\Delta, i} \overline{w^\Delta_i} / \overline{w_i}}{d_i(d_i-1)} = C^{\text{bin}}_i \frac{ \overline{w^\Delta_i}}{\overline{w_i}}
	\label{eq:app-barrat}
\end{equation}
with $n_{\Delta, i}$ the number of triangles to which node $i$ participates and $\overline{w^\Delta_i}$ the average weight associated to edges connected to $i$ and participating in a triangle.

Notice that this expression also explains why the definition from Barrat is so close to the binary clustering: for networks where weights are either rather homogeneous or where they are not strongly correlated to triangles, $\overline{w^\Delta_i}$ and $\overline{w_i}$ become very close as the number of triangles per node increases.

Using equation \ref{eq:app-barrat}, the new clustering can be defined as:

\begin{equation}
\begin{split}
	C^{B'}_i 	& = C^{bin'}_i \frac{ \overline{w^\Delta_i}'}{\overline{w_i}'} \\
				& = \frac{n_{\Delta, i}'}{d_i(d_i + 1)} \frac{\frac{n_{\Delta, i} \overline{w^\Delta_i} + (n_{\Delta, i}' - n_{\Delta, i})\epsilon}{n_{\Delta, i}'}}{\frac{d_i \overline{w_i} + \epsilon}{d_i + 1}}\\
				& = \frac{n_{\Delta, i}  \overline{w^\Delta_i}}{d_i^2 \overline{w_i}} + O(\epsilon)\\
				& = \frac{d_i - 1}{k} C^B_i + O(\epsilon)
\end{split}
\end{equation}

\subsection{Onnela}
\label{app:onnela}

As for the other definitions, the clustering from \cite{Onnela2005} can be defined as a function of the binary clustering:

\begin{equation}
	C^O_i = \frac{n_\Delta \overline{I^O_{\Delta ijk}}}{d_i(d_i-1)} = C^{\text{bin}}_i \overline{I^O_{\Delta ijk}}
\end{equation}

From this, one can define the evolution upon addition of an edge $(i, v)$ of weight $\epsilon$ as:

\begin{equation}
\begin{split}
	C^{O'}_i 	& = C^{bin'}_i \overline{I^O_{\Delta ijk}}' \\
			& = \frac{n_\Delta'}{d_i(d_i+1)} \frac{n_\Delta \overline{I^O_{\Delta ijk}} + O(\epsilon)}{n_\Delta'} \\
			& = \frac{n_\Delta}{d_i(d_i + 1)}\overline{I^O_{\Delta ijk}} + O(\epsilon) \\
			& = \frac{d_i - 1}{d_i + 1} C^O_i + O(\epsilon)
\end{split}
\end{equation}

\subsection{Directed versions of the clustering coefficients}
\label{app:directed}

To generalize the Zhang-Horvath definition of clustering~\cite{Zhang2005} for directed graphs we use the same approach as proposed in Fagiolo~\cite{Fagiolo2007}. 
These definitions are visible in Table~\ref{tab:partial-clustering_others} together with the directed definitions associated to Barrat~\cite{Barrat2004} and Onnela~\cite{Onnela2005}, respectively defined in~\citep{Clemente2018} and~\cite{Fagiolo2007}. 

\begin{table*}[t]
	\centering
	\begin{ruledtabular}
		\begin{tabular}{>{\centering\arraybackslash} m{0.08\textwidth} >{\centering\arraybackslash} m{0.18\textwidth} >{\centering\arraybackslash} m{0.19\textwidth} >{\centering\arraybackslash} m{0.275\textwidth} >{\centering\arraybackslash} m{0.09\textwidth} >{\centering\arraybackslash} m{0.15\textwidth}}
			\textbf{Mode} & $\boldsymbol{I^{O,(m)}_{\Delta,i}}$ & $\boldsymbol{I^{B,(m)}_{\Delta,i}}$ & $\boldsymbol{I^{B,(m)}_{T,i}}$ & $\boldsymbol{I_{\Delta, i}^{Z,(m)}}$ & $\boldsymbol{I_{T, i}^{Z, (m)}}$\\\hline
			Cycle & $\left(W^{\left[\frac{1}{3}\right]}\right)^3_{ii}$ & $\frac{1}{2}\left( W A^2 + \left(W A^2\right)^T \right)_{ii}$ & $\frac{1}{2}\left(s_{i, in}d_{i, out}  + s_{i, out}d_{i, in}\right) - s^B_{i,\leftrightarrow}$ & $\left(W\right)^3_{ii}$ & $s_{i, in}s_{i, out} - s^{[2]}_{i,\leftrightarrow}$\\
			Middleman & $\left(W^{\left[\frac{1}{3}\right]}W^{\left[\frac{1}{3}\right]T}W^{\left[\frac{1}{3}\right]}\right)_{ii}$ & $\frac{1}{2}\left( W A^T A + W^T A A^T\right)_{ii}$ & $\frac{1}{2}\left(s_{i, in}d_{i, out}  + s_{i, out}d_{i, in}\right) - s^B_{i,\leftrightarrow}$ & $\left(W W^{T}W \right)_{ii}$ & $s_{i, in}s_{i, out} - s^{[2]}_{i,\leftrightarrow}$\\
			Fan-in & $\left(W^{\left[\frac{1}{3}\right]T}\left(W^{\left[\frac{1}{3}\right]}\right)^2\right)_{ii}$ & $\frac{1}{2} \left(W^T(A + A^T)A\right)_{ii}$ & $s_{i, in} (s_{i, in} - 1)$ & $\left(W^{T} W W\right)_{ii}$ & $\left(s_{i, in}\right)^2 - s^{[2]}_{i, in}$\\
			Fan-out & $\left(\left(W^{\left[\frac{1}{3}\right]}\right)^2 W^{\left[\frac{1}{3}\right]T}\right)_{ii}$ & $\frac{1}{2}\left(W(A+A^T)A^T  \right)_{ii}$ & $s_{i, out} (s_{i, out} - 1)$ & $\left(W W W^{T}\right)_{ii}$ & $\left(s_{i, out}\right)^2 - s^{[2]}_{i, out}$
		\end{tabular}
	\end{ruledtabular}
	\caption{Definitions of the Barrat, Onnela, and Zhang-Horvath intensities for each partial mode pattern in directed graph.
		Column 1: pattern names; column 2: patterns illustration; column 3: Barrat triangle intensity for node $i$; column 4: Onnela triangle intensity for node $i$; column 5: Zhang--Horvath triangle intensity for node $i$; column 6: Zhang--Horvath triplet intensity for node $i$.
		The clustering coefficients associated to each mode $m$ are given by $C^{O,(m)}_i = I^{O,(m)}_{\Delta, i} / n^{(m)}_{T,i}$ for Onnela, $C^{B,(m)}_i = I_{\Delta, i}^{B,(m)}/I_{T, i}^{B,(m)}$ Barrat and by $C_i^{Z,(m)} = I_{\Delta, i}^{Z,(m)}/I_{T, i}^{Z,(m)}$ for Zhang--Horvath.
		Note that, for Barrat, the reciprocal strength has been defined in \citep{Clemente2018} as $s^B_{i, \leftrightarrow} = \sum_{i\neq j} \frac{1}{2} (w_{ij} + w_{ji})$.}
	\label{tab:partial-clustering_others}
\end{table*}

\section{Closure}
\label{app:closure}

Closure was introduced in~\cite{Yin2019} as a complementary measure of clustering for binary undirected networks.

\subsection{Undirected weighted closure}

From the Zhang--Horvath definition of clustering, closure can be generalized in a fully-weighted but non-continuous way as
\begin{equation}
	H_i^0 = \frac{\sum_{j\neq k} w_{ij} w_{jk} w_{ki}}{\sum_{j\neq k\neq i} w_{ij}w_{jk}} = \frac{W^3_{ii}}{\sum_j w_{ij} (s_j - w_{ij})}
\end{equation}
again comparing the triangle intensities to their maximum possible value if all open-walks of length two were closed into a triangle by an edge of weight 1.

Finally, it can also be defined in a continuous way as:
\begin{equation}
	H_i = \frac{\sum_{j\neq k} \sqrt[3]{w_{ij} w_{jk} w_{ki}}^2}{\sum_{j\neq k\neq i} \sqrt{w_{ij}w_{jk}}} = \frac{\left( W^{\left[ \frac{2}{3} \right]} \right)_{ii}^3}{\sum_j \left( W^{\left[ \frac{1}{2} \right]} \right)_{ij}^2 - s_i}
\end{equation}

\subsection{Directed weighted closure}

For directed networks, contrary to \cite{Yin2020}, we only consider the extension of the undirected measure to directed walks and therefore define only four variants of the directed closure, two for outgoing walks (cycle-out, CO, and fan-out, FO) and two for incoming walks (cycle-in, CI, and fan-in, FI).

We define the weighted version either directly using the weights, as in the Zhang--Horvath definition for the clustering coefficient:
\begin{equation}
	H_{i,CO}^0 = \frac{\sum_{j\neq k} w_{ij} w_{jk} w_{ki}}{\sum_{j\neq k \neq i} w_{ij} w_{jk}} = \frac{W^3_{ii}}{\sum_j w_{ij} (s_{j, out} - w_{ij})}
\end{equation}

\begin{equation}
	H_{i,CI}^0 = \frac{\sum_{j\neq k} w_{kj} w_{ji} w_{ik}}{\sum_{j\neq k \neq i} w_{kj} w_{ji}} = \frac{\left(W^T\right)^3_{ii}}{\sum_j w_{ji} (s_{j, in} - w_{ji})}
\end{equation}

\begin{equation}
	H_{i,FO}^0 = \frac{\sum_{j\neq k} w_{ij} w_{jk} w_{ik}}{\sum_{j\neq k \neq i} w_{ij} w_{jk}} = \frac{\left(W^2 W^T\right)_{ii}}{\sum_j w_{ij} (s_{j, out} - w_{ij})}
\end{equation}

\begin{equation}
	H_{i,FI}^0 = \frac{\sum_{j\neq k} w_{kj} w_{ji} w_{ki}}{\sum_{j\neq k \neq i} w_{kj} w_{ji}} = \frac{\left(W^T W^2\right)_{ii}}{\sum_j w_{ji} (s_{j, in} - w_{ji})}
\end{equation}

Or via the continuous definition:
\begin{equation}
	H_{i,CO}^c	= \frac{\sum_{j\neq k} \sqrt[3]{w_{ij} w_{jk} w_{ki}}^2}{\sum_{j\neq k \neq i} \sqrt{w_{ij} w_{jk}}}
				= \frac{\left( W^{\left[ \frac{2}{3} \right]} \right)_{ii}^3}{\sum_j \left( W^{\left[ \frac{1}{2} \right]} \right)_{ij}^2 - s_{i, out}}
\end{equation}

\begin{equation}
	H_{i,CI}^c = \frac{\sum_{j\neq k} \sqrt[3]{w_{kj} w_{ji} w_{ik}}^2}{\sum_{j\neq k \neq i} \sqrt{w_{kj} w_{ji}}} = \frac{\left( W^{\left[ \frac{2}{3} \right],T} \right)_{ii}^3}{\sum_j \left( W^{\left[ \frac{1}{2},T \right]} \right)_{ij}^2 - s_{i, in}}
\end{equation}

\begin{equation}
	H_{i,FO}^c	= \frac{\sum_{j\neq k} \sqrt[3]{w_{ij} w_{jk} w_{ki}}^2}{\sum_{j\neq k \neq i} \sqrt{w_{ij} w_{jk}}}
				= \frac{\left( {W^{\left[ \frac{2}{3} \right]}}^2 W^{\left[ \frac{2}{3} \right],T} \right)_{ii}}{\sum_j \left( W^{\left[ \frac{1}{2} \right]} \right)_{ij}^2 - s_{i, out}}
\end{equation}

\begin{equation}
	H_{i,FI}^c = \frac{\sum_{j\neq k} \sqrt[3]{w_{kj} w_{ji} w_{ki}}^2}{\sum_{j\neq k \neq i} \sqrt{w_{kj} w_{ji}}} = \frac{\left( W^{\left[ \frac{2}{3} \right],T} {W^{\left[ \frac{2}{3} \right]}}^2 \right)_{ii}}{\sum_j \left( W^{\left[ \frac{1}{2},T \right]} \right)_{ij}^2 - s_{i, in}}
\end{equation}

\section{Network generation algorithms}

All networks were generated using the \href{https://nngt.readthedocs.io/}{NNGT} library \cite{NNGT}.

\subsection{Core-periphery network}
\label{subsec:core-periphery}

The core-periphery network on Figure~\ref{fig:core-periph} contains:
\begin{itemize}
	\item 1 central core node (CCN)
	\item 10 outer-core nodes (OCNs)
	\item 20 periphery nodes (PNs)
\end{itemize}

The nodes are connected as follow:
\begin{itemize}
	\item the 10 OCNs for a circular graph with fully reciprocal connections to their 4 nearest-neighbors,
	\item the OCNs all connect to the CCN with reciprocal connections.
\end{itemize}
The weights associated to these connections are drawn from $U(5, 10)$.
The connections with the PNs are as follow:
\begin{itemize}
	\item each OCN receive one connection from every other PN, starting with the first or second PN depending on the OCN's evenness,
	\item the OCN reciprocate the connections with probability 0.5
	\item the PNs are connected among themselves following an Erd\H{o}s--Renyi scheme of density 0.05.
\end{itemize}
Weights associated to connections involving PNs are drawn from $U(0.05, 0.5)$.

\subsection{Watts--Strogatz}

The original Watts--Strogatz network \cite{Watts1998} consists of a regular lattice basis (characterized by a coordination number $k$) that is then modified, rewiring each edge with a probability $p$.
For directed networks, we used a generalization of that method implemented in NNGT which is strictly equivalent except for the fact that edges are now directed:
\begin{enumerate}
	\item start from a directed regular lattice with coordination number $k$ and reciprocity $r$ (taken as 1 in this paper),
	\item rewire each edge with probability $p$.
\end{enumerate}

The original lattice $L(N, k, r)$ has $\frac{1}{2}Nk (1 + r)$ edges, leading to the limit cases
\begin{itemize}
	\item $\frac{1}{2}Nk$ edges if $r = 0$, like the undirected lattice,
	\item $Nk$ edges if $r = 1$, with all connections being reciprocal.
\end{itemize}

In the network used on Figure \ref{fig:spurious_edges}, we used a coordination number $k = 20$ and a rewiring probability $p = 0.03$.

\section{Real-world networks}

\subsection{Mouse mesoscale connectome}

The network was obtained from \cite{Oh2014} and contains 426 nodes corresponding to brain regions of intermediate scale connected by 65,465 edges.
It is a symmetric version of the original network that separates nodes from both hemispheres.
Each node is associated a ``name'' property corresponding of a abbreviated denomination for the corresponding brain region, as well as a suffix (left, or right) corresponding to the hemisphere.
Edges are associated three attributes: a ``weight'', a ``pvalue'', and a ``distance'' (corresponding to the Euclidean distance from the source to the target node).

\vspace*{1ex}

\begin{table}[H]
	\begin{ruledtabular}
	\begin{tabular}{r c c c c c}
	& \textbf{Mean} & \textbf{STD} & \textbf{Median} & \textbf{Min} & \textbf{Max}\\
	\hline
	In-degree & 153.7 & 25.6 & 152 & 104 & 223\\
	Out-degree & 153.7 & 77.6 & 151 & 14 & 357\\
	$CC^{bin}_{tot}$ & 0.43 & 0.02 & 0.43 & 0.38 & 0.49\\
	Weights & 0.076 & 0.36 & 0.038 & 3.9$\cdot 10^{-16}$ & 20.4 
	\end{tabular}
	\end{ruledtabular}
	\caption{Some characteristic properties of the mesoscale mouse brain.}
\end{table}

\begin{figure}[h]
	\includegraphics[width=\linewidth]{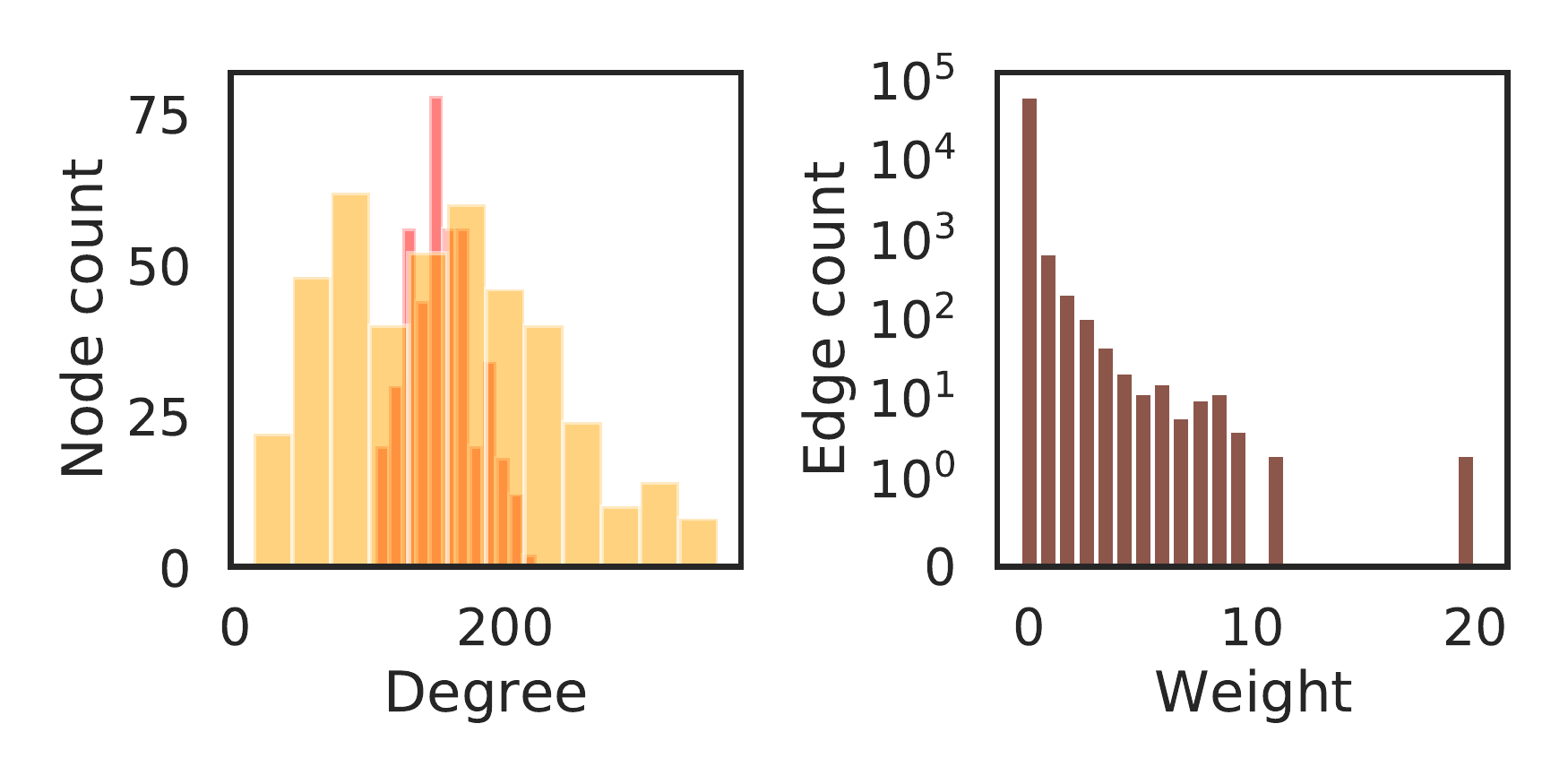}
	\caption{Distribution of in/out-degrees (respectively in orange/yellow) and weights in the mouse connectome.}
\end{figure}

\subsection{Fediverse mesoscale network}
\label{app:fediverse}

The network was obtained from~\cite{Zignani2019}\footnote{data available at \href{https://dataverse.mpi-sws.org/dataset.xhtml?persistentId=doi:10.5072/FK2/AMYZGS}{https://dataverse.mpi-sws.org/dataset.xhtml?persistentId=doi:10.5072/FK2/AMYZGS}}.
It contains 3,825 nodes representing servers (instances) that are connected via 81,371 edges.
Each node is associated two attributes: a ``name'', corresponding to the server domain, and a ``size'', $S_i$, corresponding to the number of users registered on that server.
Edges are associated four attributes: ``num\_follows'', defined a $F_{ij}$, the number of followers from $i$ to $j$, ``follow/size'' defined as $F_{ij} / S_i$, a ``weight'' given by $w_{ij} = F_{ij} / F_i$, the ratio between the number of followers from $i$ to $j$ divided by the total number of followers from $i$, and a ``distance'' giving the Euclidean distance between the two servers on the latitude/longitude plane.

\begin{table}[h]
	\begin{ruledtabular}
	\begin{tabular}{r c c c c c}
	& \textbf{Mean} & \textbf{STD} & \textbf{Median} & \textbf{Min} & \textbf{Max}\\
	\hline
	In-degree & 21.3 & 82.6 & 4 & 0 & 2271\\
	Out-degree & 21.3 & 77.8 & 5 & 0 & 2038\\
	$CC^{bin}_{tot}$ & 0.68 & 0.37 & 0.86 & 0 & 1\\
	Weights & 0.045 & 0.14 & 0.0029 & 6.0$\cdot 10^{-7}$ & 1 
	\end{tabular}
	\end{ruledtabular}
	\caption{Some characteristic properties of the Fediverse mesoscale network.}
\end{table}

\begin{figure}[h]
	\includegraphics[width=\linewidth]{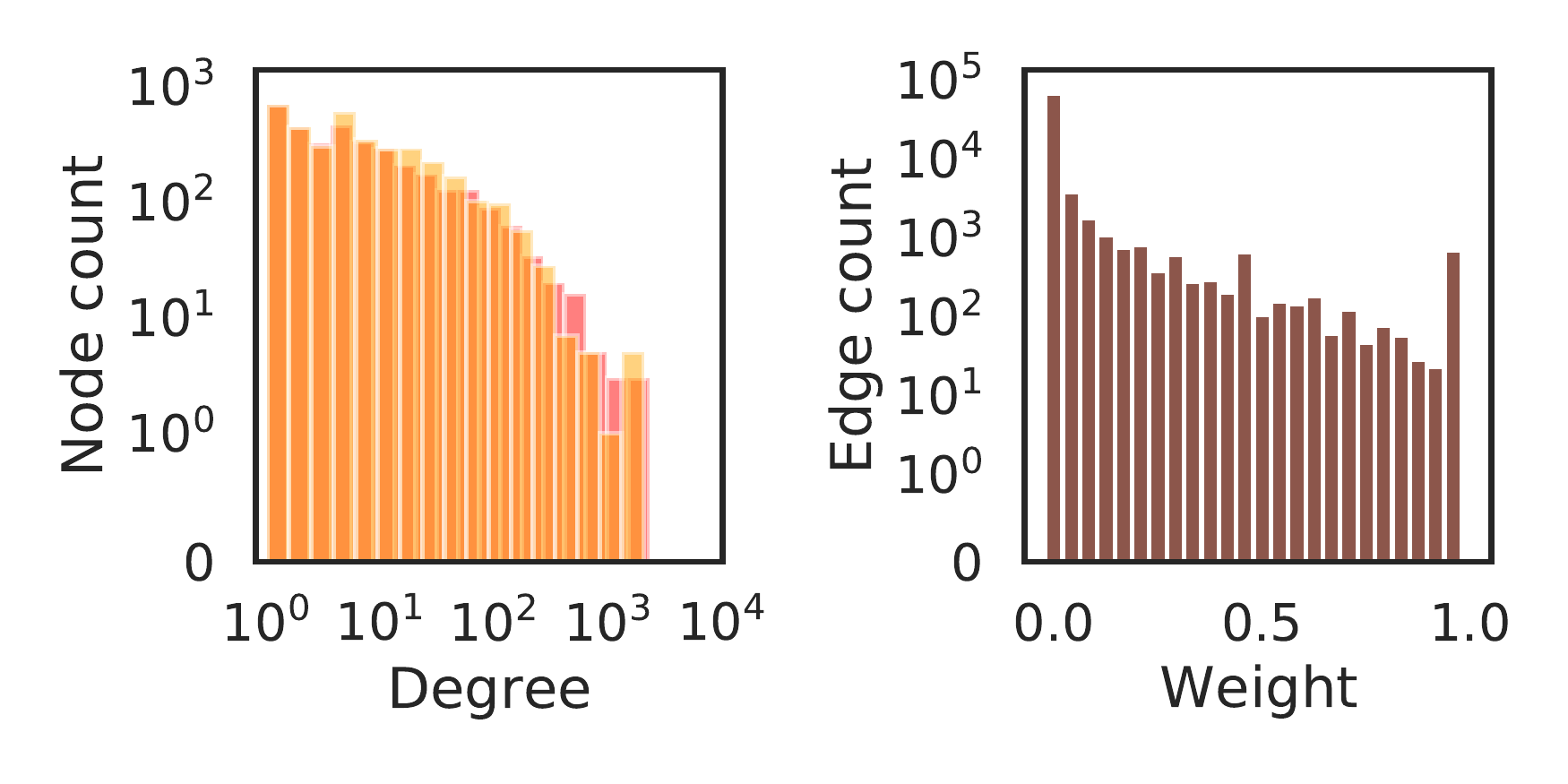}
	\caption{Distribution of in/out-degrees (respectively in orange/yellow) and weights in the Fediverse instances.}
\end{figure}

\subsection{Closure in the shuffled networks}
\label{app:closure-shuffled}

\begin{figure}[b]
	\includegraphics[width=\textwidth]{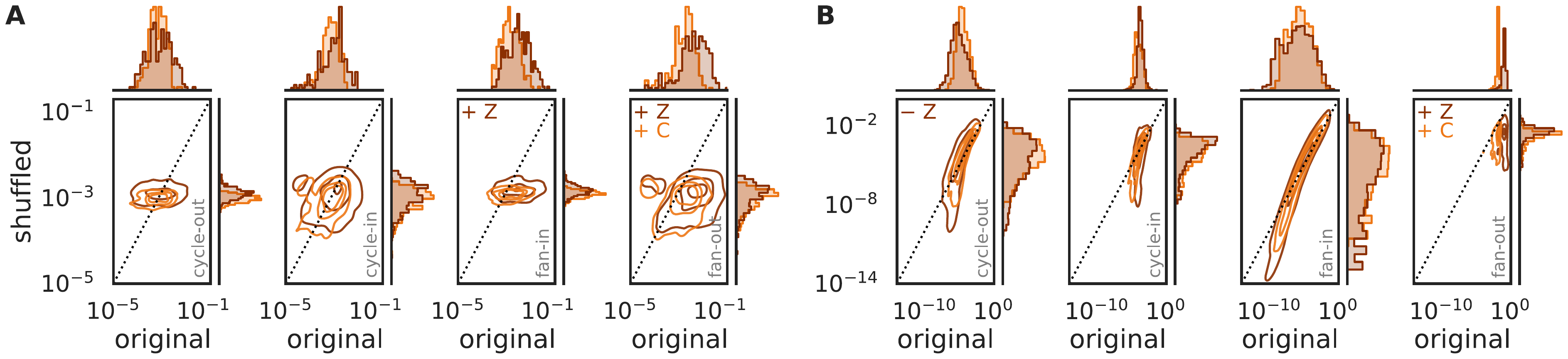}
	\caption{
		Different structures of local closure in the mouse brain and Fediverse.
		The panels compare the local closure of directed patterns for the mouse brain (A) and Fediverse (B).
		The original values are compared to the averages over an ensemble of graphs with the same adjacency matrix but shuffled edge weights. For mouse: 10 uniformly shuffled networks, for Fediverse: 200 shuffling across all out-going edges of a each node (to preserve out-strength normalization).
		\textbf{A}. The local closure also shows an increase for patterns promoting redundancy of information flow in the mouse brain (fan-in/fan-out motifs are higher in the original graph).
		\textbf{B}. For the Fediverse, only fan-out is significantly stronger than in the random graphs.
		Patters where the original values are significantly greater (smaller) than the randomized ones are marked by the $+$ (resp. $-$) and initial of the method (brown, Z for Zhang; orange, C for continuous; yellow, O for Onnela).
		The original closure is considered to be significantly higher (resp. lower) if 75\% of the points where above (resp. below) the dotted identity line.
		Additional * denotes one-sided fractions greater than 95\%.}
\label{fig:kde-closure}
\end{figure}

The results obtained for the local closure confirm those obtained using the clustering coefficient for the mouse brain and Fediverse networks, as show on Figure~\ref{fig:kde-closure}.
Namely, we see an overexpression of fan-in and fan-out patterns for the mouse and a discrepancy between those two patterns for the Fediverse.

\subsection{Networks with a high number of single-node triangles}
\label{app:single-tr}

The presence of nodes participating to a single triangle is frequently in very sparse networks such as collaboration or gene/protein interaction networks.

Table \ref{tab:1-tr} details several such networks:
\begin{description}
	\item[NetScience] co-authorship in the network science community\footnote{\href{https://networkrepository.com/netscience.php}{https://networkrepository.com/netscience.php}}
	\item[CompGeo] collaborations in computational geometry\footnote{\href{http://vlado.fmf.uni-lj.si/pub/networks/data/collab/geom.htm}{http://vlado.fmf.uni-lj.si/pub/networks/data/collab/geom.htm}}
	\item[CE-CX] a graph of gene associations inferred from co-expression pattern of two genes (based on high-dimensional gene expression data) for \textit{C. elegans}
	\item[CE-HT] gene associations inferred from high-throughput protein-protein interactions for \textit{C. elegans}
	\item[HS-HT] gene associations inferred from high-throughput protein-protein interactions for \textit{Homo sapiens}
	\item[SC-HT] gene associations inferred from high-throughput protein-protein interactions for \textit{S. cerevisiae}
\end{description}

Gene functional association networks \cite{Cho2014} were downloaded from \href{https://networkrepository.com/bio.php}{https://networkrepository.com/bio.php}.

\begin{table}[h]
	\begin{ruledtabular}
	\begin{tabular}{r c c c}
		\textbf{Network} & \textbf{N} & \textbf{E} & \textbf{1-T nodes (\%)}\\\hline
		NetScience	& 1,589		& 2,742		& 23	\\
		CompGeo 		& 7,343		& 11,898		& 21	\\
		CE-CX 		& 15,229		& 24,5952	& 8\\
		CE-HT		& 2,617		& 2,985		& 2\\
		HS-HT		& 2,570		& 13,691		& 7\\
		SC-HT		& 2,084		& 63,027		& 5
	\end{tabular}
	\end{ruledtabular}
	\caption{Number of nodes (N), edges (E), and fraction of single-triangle (1-T) nodes in several collaboration and gene association networks.}
	\label{tab:1-tr}
\end{table}

\onecolumngrid

\end{document}